\documentclass[preprint,12pt,authoryear]{elsarticle}
\usepackage{amssymb}
\usepackage{amsmath}
\usepackage[hyphens]{url}
\usepackage{hyperref}
\usepackage{subfig}
\usepackage{pdflscape}
\usepackage[ruled,vlined]{algorithm2e}
\SetKw{KwBy}{by}

\journal{Astronomy and Computing}

\begin{document}

\begin{frontmatter}

\title{FITrig: A High-Performance Detection Technique for Efficient Ultra-Long-Period Pulsars} 

\author{Xiaotong Li, Karel Ad\'{a}mek, Wesley Armour} 

\affiliation{organization={Oxford e-Research Centre, Department of Engineering Science, University of Oxford},
            addressline={7 Keble Rd}, 
            city={Oxford},
            postcode={OX1 3QG}, 
            country={UK}}
\begin{abstract}
Ultra-long-period (ULP) pulsars, a newly identified class of celestial transients, offer unique insights into astrophysics, though very few have been detected to date. In radio astronomy, most time-domain detection methods cannot find these pulsars, and current image-based detection approaches still face challenges, including low sensitivity, high false positive rate, and low computational efficiency. In this article, we develop Fast Imaging Trigger (FITrig), a GPU-accelerated, statistics-based method for ULP pulsar detection and localisation. FITrig includes two complementary approaches --- an image domain and an image-frequency domain strategy. FITrig offers advantages by increasing sensitivity to faint pulsars, suppressing false positives (from noise, processing artefacts, or steady sources), and improving search efficiency in large-scale wide-field images. Compared to the state-of-the-art source finder SOFIA 2, FITrig increases the detection speed by 4.3 times for large images ($50\mathrm{K} \times 50\mathrm{K}$ pixels) and reduces false positives by up to 858.8 times (at 6$\sigma$ significance) for the image domain branch, while the image-frequency domain branch suppresses false positives even further. FITrig maintains the capability to detect pulsars that are 20 times fainter than surrounding steady features, even under critical Nyquist sampling conditions. In this article, the performance of FITrig is demonstrated using both real-world data (MeerKAT observations of PSR J0901-4046) and simulated datasets based on MeerKAT and SKA Array Assembly (AA) 2 telescope configurations. With its real-time processing capabilities and scalability, FITrig is a promising tool for next-generation telescopes, such as the SKA, with the potential to uncover hidden ULP pulsars.
\end{abstract}

\begin{highlights}
    \item \textbf{tLISI --- a novel transient-oriented IQA index:} A statistically-grounded IQA metric specially designed to highlight pulsar-specific variations through probability theory, enabling more reliable transient detection.
    \item \textbf{Elimination of deconvolution:} FITrig completely removes the need for deconvolution in the detection process while maintaining accuracy, significantly reducing computational overhead.
    \item \textbf{Dual-domain detection framework:} Two branches of FITrig (image domain and image-frequency domain approaches) are developed, significantly improving the capability to detect faint ULP pulsars.
    \item \textbf{GPU-powered real-time processing:} FITrig is highly parallelised with the GPU implementation, enabling real-time processing and scalability for large datasets.
\end{highlights}


\end{frontmatter}

\section{Introduction}

Searching for pulsars \citep{Pulsar} is a popular topic in radio astronomy since detecting and observing their radio emissions allows astronomers to test theories such as general relativity \citep{relativity,relativity1} and to study astrophysical processes governed by magnetohydrodynamics (MHD) \citep{MHD,MHD1,MHD2}. Ultra-Long-Period (ULP) Pulsars, such as PSR J0901-4046 with its exceptionally long 76-second rotation period \citep{realpul2}, are a unique class of pulsars characterised by unusually slow spins. ULP pulsars could reveal hidden neutron star populations and improve gravitational wave studies, while their emission properties also challenge traditional models of neutron star physics \citep{ulf}.

However, ULP pulsars introduce significant detection challenges, as their extremely slow spin rates produce faint, widely spaced radio pulses that are often masked by noise. Even harder, their long-period emission and transient nature demand prolonged observations to reliably pinpoint true signals from false positives. These challenges explain why most pulsar detection techniques, which operate primarily in the time-frequency domain \citep{Pulsar,fdas1,fdas2}, struggle to identify ULP pulsars. 

\begin{figure}[!t]
    \centering
    \includegraphics[width=\textwidth]{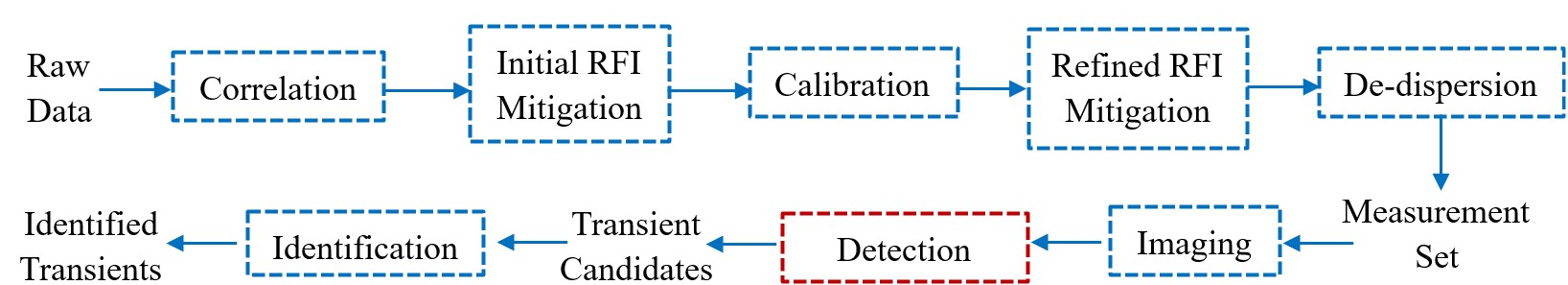}
\caption{Schematic representation of an image-based radio transient detection workflow, where the highlighted ``Detection'' component indicates the primary focus of this article.
\label{addflow}}
\end{figure}

Transitioning pulsar search from the time-frequency domain to the image domain enables the detection of ULP pulsars because their long-period signals form coherent patterns (e.g., repeating bright spots) that are more easily distinguished from noise than widely spaced peaks in traditional periodicity searches. Fast Imaging (FI) reconstructs images from visibilities collected over short timescales and identifies transient candidates from these images (referred to as ``snapshots'') through statistical selection \citep{srcdtc}. Recent astrophysical results from image-based transient searches include the detection of binary white dwarf pulsars \citep{realpul1} and ULP neutron stars \citep{realpul2} in radio bands. This area has motivated detection tools such as \textit{realfast} \citep{realfast} for the Karl G. Jansky Very Large Array \footnote{\url{https://public.nrao.edu/telescopes/vla/}} (VLA; \citealt{VLAapjs}). Table \ref{procon} presents a comprehensive analysis that compares the benefits and limitations of pulsar detection in the image domain against the time domain. An example workflow for image-based radio pulsar detection is illustrated in Fig. \ref{addflow} \citep{RAbook,RAbook2}.
\begin{table}[t]
\centering
\begin{tabular}{p{0.2\textwidth} p{0.35\textwidth} p{0.35\textwidth}}
\hline
Aspect & Time Domain Detection & Image Domain Detection\\
\hline
Event types detected & Ideal for high-frequency pulsars. & Better for low-frequency pulsars.\\
Temporal resolution & Excellent temporal resolution. & Limited by image integration time.\\
Spatial resolution & Poor spatial resolution. & High spatial resolution.\\
Computational demand & Lower; processing focused on time-series analysis. & High; imaging involves Fourier transforms and deconvolution.\\
Data volume & Relatively smaller data volumes. & Larger data volumes.\\
\hline
\end{tabular}
\caption{Comparisons of time-domain vs. image-domain pulsar detection.}\label{procon}
\end{table}

In the FI workflow, after creating wide-field images for each time slot, pulsars are localised in the resulting snapshots using source-finding techniques, such as the SOurce FInding Application (SOFIA) 2 \citep{sofia2}, SFIND \citep{sfind}, AEGEAN \citep{sourcefinder1}, and Python Blob Detector and Source Finder \footnote{\url{https://www.ascl.net/1502.007}} (PyBDSF). Among them is SOFIA 2, a parallel code that utilises OpenMP for multi-threading in time-critical parts of the implementation. In it, the input image is smoothed across many spatial and spectral scales to pinpoint notable emissions. This algorithm is known as the Smooth and Clip (S + C) finder \citep{scfinder}.

However, challenges remain in current source finders for ULP pulsar detection:
\begin{itemize}
    \item \textbf{Sensitivity:} The detection of pulsars, particularly for ULP pulsars, is often limited by their low signal-to-noise ratio (SNR) in radio astronomical images. 
    \item \textbf{False Positive Rate:} Robust separation of true pulsar signals from artefacts is critical to avoid misclassifying false positives as real ULP pulsars.
    \item \textbf{Computational Efficiency:} The growing volume of data from next-generation telescopes demands more efficient pulsar finding algorithms. A computationally optimised ULP detection technique is essential for real-time analysis, enabling timely follow-up observations of candidate pulsars.
\end{itemize}

To address these challenges, we develop a new detection technique for ULP pulsars, called Fast Imaging Trigger (FITrig). FITrig achieves high sensitivity through a target-sensitive image assessment method, tLISI (transient-oriented Low-Information Similarity Index), proposed in this work. False positives are minimised by a statistic-based auto-selection process integrated into the triggering workflow. In addition, high computational efficiency is achieved through the GPU-accelerated design of FITrig, which leverages its highly parallel architecture to rapidly process large datasets and enable timely identification of candidate pulsars.

To support clarity throughout this article, some key terms are briefly explained here. A ``celestial transient'', or simply ``transient'', refers to a radio source whose emission varies noticeably over time in observations. A ``steady source'', in contrast, denotes a radio source whose emission remains largely constant, showing no significant variability during observations. The ``dirty beam'' is the point spread function (PSF) of the telescope array resulting from its incomplete \textit{uv}-coverage; we use the term ``dirty beam'' throughout. Finally, ``steady features'' refer to the structures formed by steady sources together with the dirty beams convolved on them. For reference, a table summarising the technical terms and their abbreviations used in this article is provided in Table \ref{abbrev}.

The methodology of FITrig is proposed in Section \ref{section2}. Its performance is demonstrated by the experimental results in Section \ref{section4}, starting with the performance of FITrig on the real dataset of PSR J0901-4046 measured by MeerKAT \citep{MKAT1,MKAT2,MKAT3}. We then test FITrig on more challenging cases using simulated datasets covering various pulsar brightness levels and frequencies, based on the telescope layouts of MeerKAT and Square Kilometre Array (SKA; \citealt{SKA1}) AA2 \footnote{\url{https://www.skao.int/en/news/174/green-light-given-construction-worlds-largest-radio-telescope-arrays}}. This further illustrates its performance in terms of computational efficiency (Section \ref{CompEffi}), sensitivity (Section \ref{413}), and false positive rate (Section \ref{sec41}). Section \ref{seccon} presents the conclusions.

\section{Method}
\label{section2}

\subsection{Problem Analysis}
\label{section21}

In radio interferometric imaging, the relationship between visibilities and the received sky brightness distribution (SBD) \citep{RAbook} is given by
\begin{equation}
V\left( {u,v,w} \right) = {\iint{I\left( {l,m} \right)e^{- i2\pi{({ul + vm + w{({\sqrt{1 - l^{2} - m^{2}} - 1})}})}}dldm}} ,
\label{vis}
\end{equation}
where $V$ represents the observed visibilities, $I$ represents the received SBD, and $l,m,n$ and $u,v,w$ are coordinates in the spatial and frequency domains, respectively. The image obtained by inverse Fourier transforming the observed visibilities is called the dirty image, where celestial sources appear convolved with the dirty beam.

In image-based pulsar detection, the influence of the dirty beam is mitigated through either deconvolution techniques \citep{realpul1} such as CLEAN \citep{CLEAN1} and Maximum Entropy Method (MEM; \citealt{MEM10}), or by computing difference images from time-adjacent snapshots. While deconvolved images yield higher detection accuracy, they require significantly more processing time. Conversely, difference images offer a faster alternative but at the cost of increased false positives and reduced sensitivity to fainter pulsars as a result of their higher noise level compared to deconvolved images.

From the perspective of computational efficiency, the latter is a better choice. However, there are some issues remaining with image-based ULP pulsar searches.
\begin{itemize}
    \item \textbf{Localised Search Regions:} When conducting wide-field transient searches that prioritise large-area coverage, pulsars often occupy only a small proportion of the image, making whole-image searches computationally wasteful.
    \item \textbf{Sparse Temporal Appearance:} ULP pulsars appear in just a few snapshots within time-sequential difference images. Blindly searching every difference frame wastes resources, especially for the large volume of data collected by next-generation telescopes.
    \item \textbf{Dynamic ``Steady'' Features:} Ideally, difference images would contain only the pulsar signal convolved with the dirty beam, as steady sources and noise would cancel out when subtracting adjacent snapshots. In reality, the dirty beam changes with Earth's rotation and steady sources fluctuate in time, resulting in visual artefacts due to partial cancellation of the dirty beam in the difference image. These artefacts, when bright enough, dominate the difference image and obscure the pulsar signal.
\end{itemize}

To address these issues, we propose a method, FITrig, tailored for detecting ULP pulsars in difference images from dirty snapshots.

\subsection{Transient-oriented Low-Information Similarity Index (tLISI)}

In order to avoid whole-image searches, FITrig is designed to start by localising potential regions that contain pulsars within the SBD. Intensity-sensitive Image Quality Assessment (IQA) methods have been developed to pinpoint potential regions containing transients \citep{adass}, allowing an efficient search of the entire image. In this context, while these approaches are conventionally termed IQAs, they function not as quality assessors, but rather as detectors. Among them, Low-Information Similarity Index (LISI; \citealt{Wider}) is suitable for assessing images with point-like sources, while augmented LISI (augLISI; \citealt{iqara}) is better for assessing images with extended structures. Aiming to process radio astronomical images, both LISI and augLISI are designed to be more sensitive to high-intensity differences in the two input images being compared, but less sensitive to low-intensity differences.

However, in the detection of ULP pulsars, high-sensitivity detection requires the detector to find fainter transients that may be obscured by dynamic steady features (as analysed in Section \ref{section21}) and noise. If an IQA method is less sensitive to low-intensity differences, those faint transients would be ignored. Therefore, in this article, we propose transient-oriented Low-Information Similarity Index (tLISI), a new IQA method for detecting ULP pulsars in radio astronomical images.

\subsubsection{ULP Pulsar Detection in Consecutive Images}
\label{sec221}

\begin{figure}[!t]
\centering
\subfloat{}{\includegraphics[width=\textwidth]{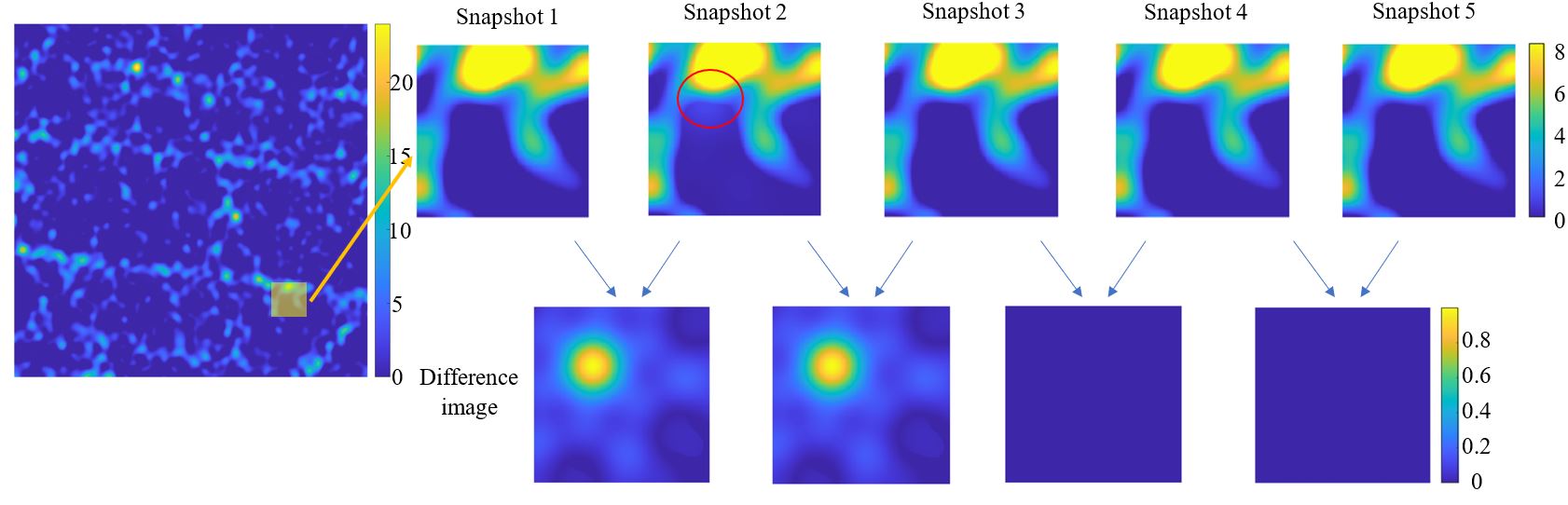}\caption*{(a)}}
\subfloat{}{\includegraphics[width=\textwidth]{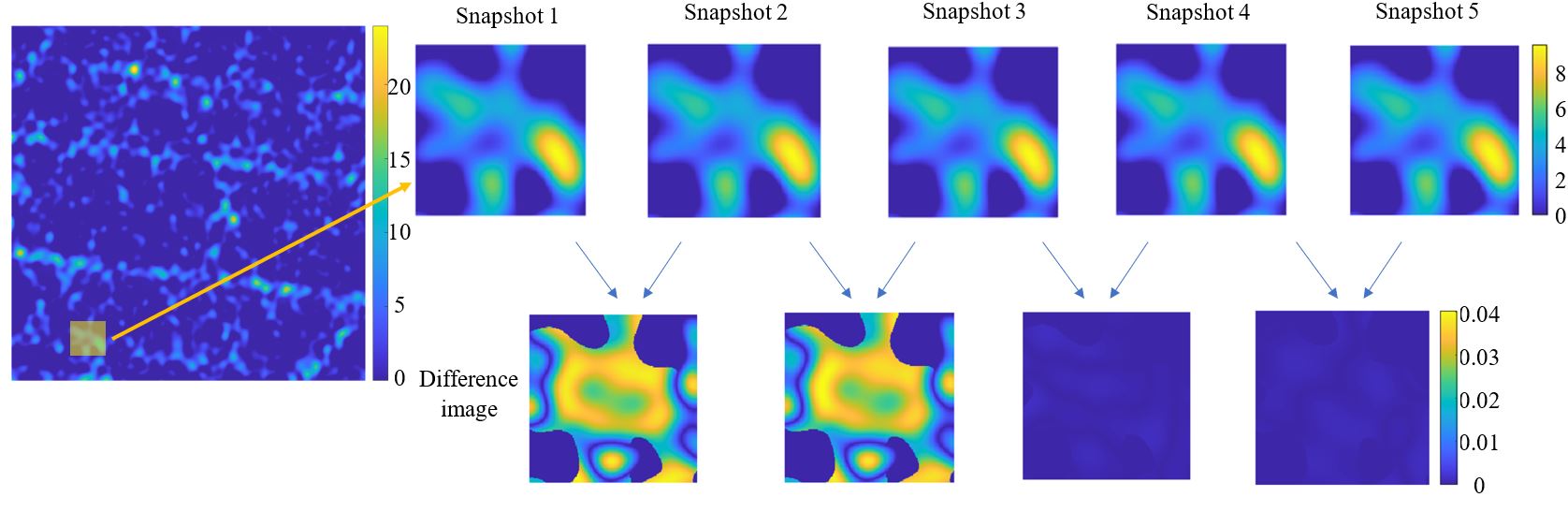}\caption*{(b)}}
\caption{Example snapshots and their difference images reconstructed from a simulated dataset generated by OSKAR using VLA D array layout.
\label{diffmodel}}
\end{figure}

As analysed in Section \ref{section21}, FITrig (starting from tLISI) is designed to operate on the difference images between adjacent snapshots, as shown in Fig. \ref{diffmodel}. The example figures were simulated by Oxford's Square Kilometre Array Radio-telescope simulator (OSKAR; \citealt{OSKAR1}). Figure \ref{diffmodel} (a) and (b) show two crops from time adjacent snapshots, with the pulsar present only in Snapshot 2 of crop (a), highlighted by the red circle. In (b), the first two difference images show shadowed shapes, which are parts of the side lobe of the convolved dirty beam, but their intensities are much lower than those of the pulsar (as shown in the first two difference images in (a)), as expected.

\begin{figure}[!t]
    \centering
    \includegraphics[width=\textwidth]{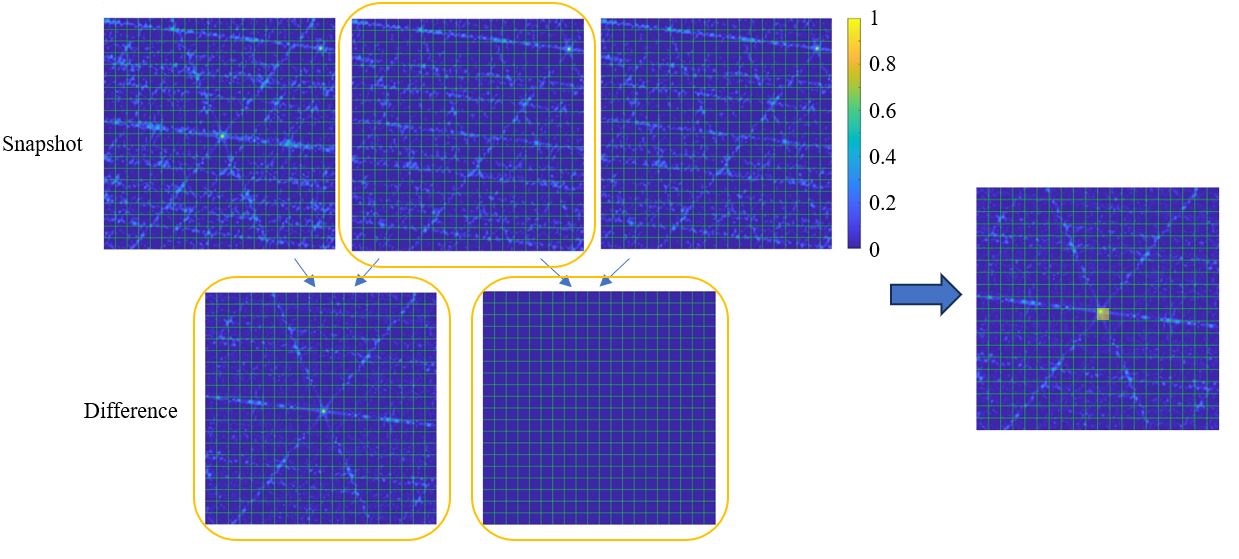}
\caption{Three-snapshot unit in the detection model. These images are reconstructed from a simulated dataset generated by OSKAR using the VLA D telescope layout. Suppose the index of the top-left tile is (1,1). In this example, the pulsar is primarily located in tile (11,11). The colour bar indicates the intensity scales, which are normalised in this example for illustration purposes. In the right figure, the yellow-shadowed tile indicates the tile selected by tLISI, which will be passed to the next step for further localisation.
\label{tiles}}
\end{figure}

Typically, a ULP pulsar has a period longer than twice the sampling period of the telescope. Given their sparse temporal appearance, a detection model is built on three-snapshot units, where the ULP pulsar appears in only one snapshot, as shown in Fig. \ref{tiles}. The three highlighted images represent the input for tLISI when comparing the two difference images. Both snapshots and difference images are divided into tiles to localise potential region and enable parallel computing. All tiles at the same position in the time-sequential images need to have the same shape, whereas the tile shape can be regular or irregular. To reduce computational costs, it is preferable to use square or rectangular tiles. The choice of tile size affects the sensitivity of pulsar detection, which will be discussed in Section \ref{section4}.

The tLISI metric is designed such that its value is higher when both difference images include a pulsar, or neither image contains a pulsar. On the other hand, the tLISI value is lower when one difference image includes a pulsar while the other does not. In our design, tLISI ($\mathrm{tLISI} \in [0,1]$) equals 1 when the two difference images are exactly the same and 0 when the two difference images are very different. Designed for parallel computing efficiency, the tLISI metric processes each tile independently.
 
\subsubsection{Side Lobes of Dirty Beam}
\label{sec222}

While pulsar snapshots (images containing the pulsar) can be identified by searching for difference images that include the dirty beam convolved with the pulsar, tLISI must accurately localise ``pulsar tiles'' (those containing the true pulsar signal) while rejecting tiles that only contain side-lobe artefacts from the dirty beam. In other words, tLISI needs to be more sensitive to tiles with larger intensity differences between adjacent snapshots. However, what is the proper metric to define ``intensity differences''?

\begin{figure}[!t]
\centering
    \includegraphics[width=4.5in]{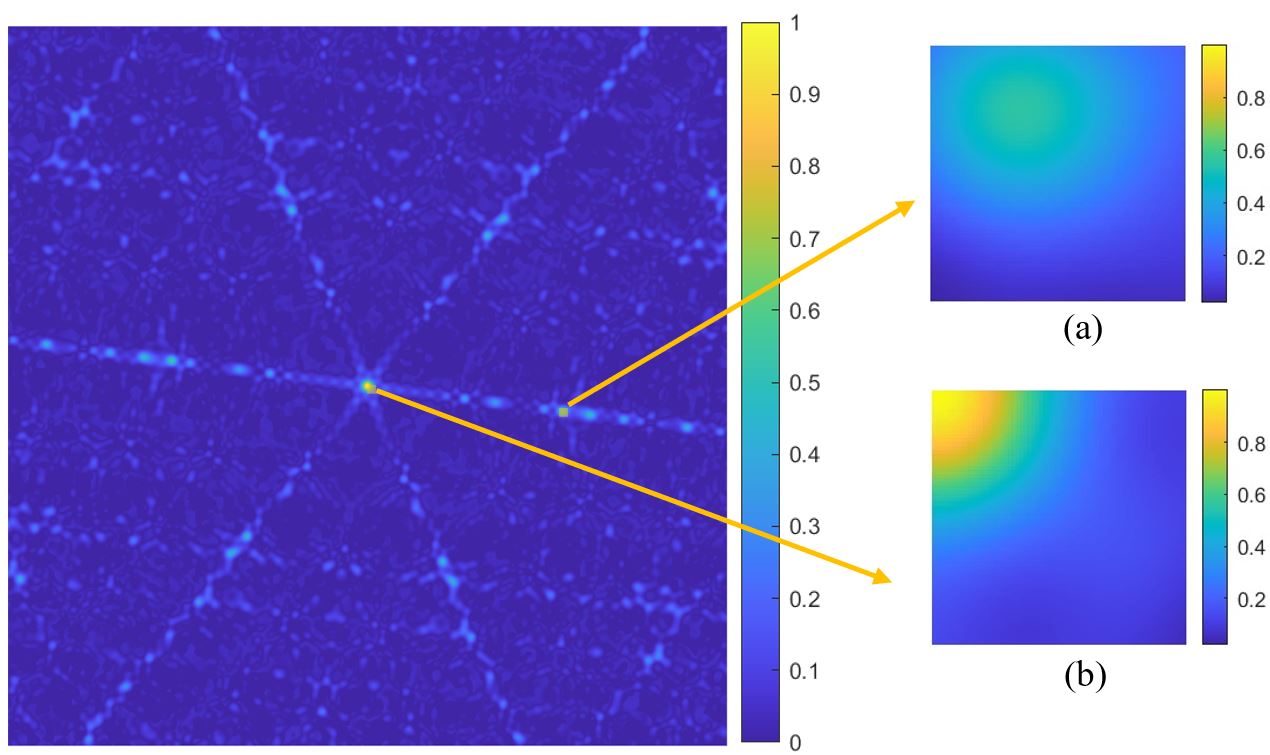}
\caption{An example of tiles in a difference image showing the side lobe effect of dirty beam: tile (a), which contains no pulsar, shows a larger sum of intensities due to the side lobes of the dirty beam convolved with the pulsar, compared to tile (b), which does contain a pulsar. This example is simulated using the VLA telescope layout.
\label{affectsum}}
\end{figure}

When the dirty beam's side lobes extend over a wide area and the main lobe is narrow, ``non-pulsar tiles'' (those without any pulsar) may still exhibit a high sum of pixel intensities, potentially even greater than that of the pulsar tile. For example, Fig. \ref{affectsum} illustrates this effect: a non-pulsar tile (Fig. \ref{affectsum} (a)) can show a significant total pixel intensity from side lobes, even though its maximum pixel intensity remains lower than that of the true pulsar tile (Fig. \ref{affectsum} (b)).

Therefore, to minimise the false positive rate, tLISI should consider both the sum of pixel intensities and the maximum intensity within a tile, ensuring robust discrimination between pulsar tiles and side lobe artefacts.

\subsubsection{Dynamic Steady Features Mitigation}
\label{sec223}

\begin{figure}[!t]
\subfloat{}{\includegraphics[width=\textwidth]{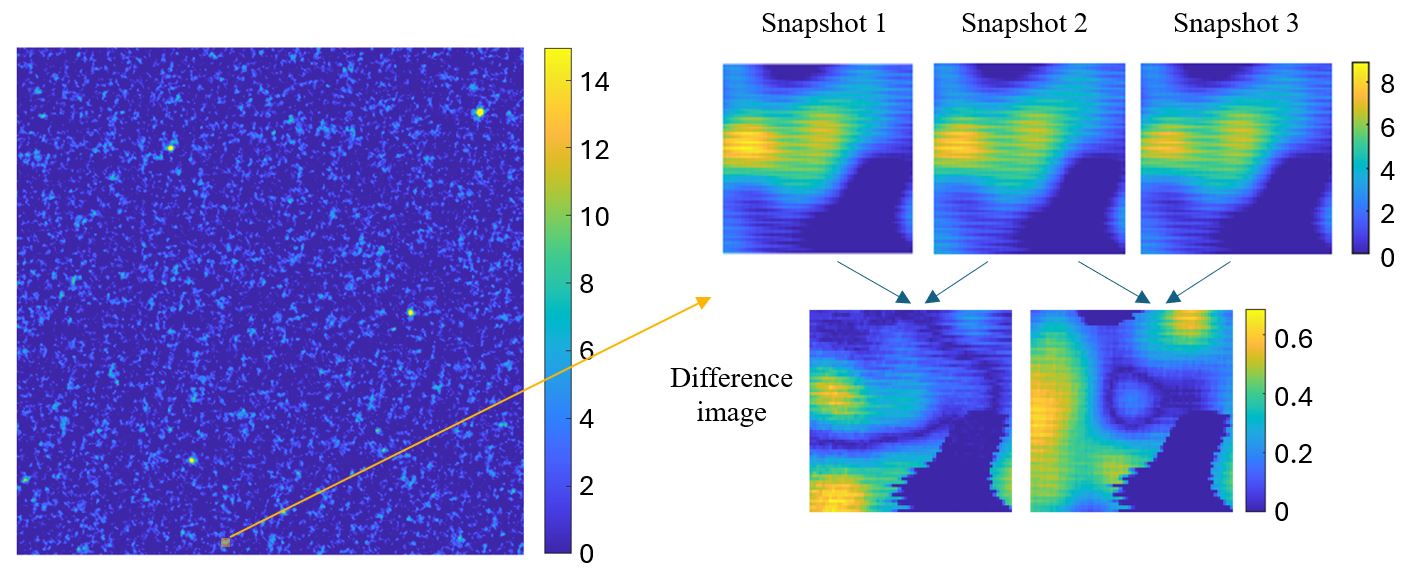}\caption*{(a)}}
\subfloat{}{\includegraphics[width=\textwidth]{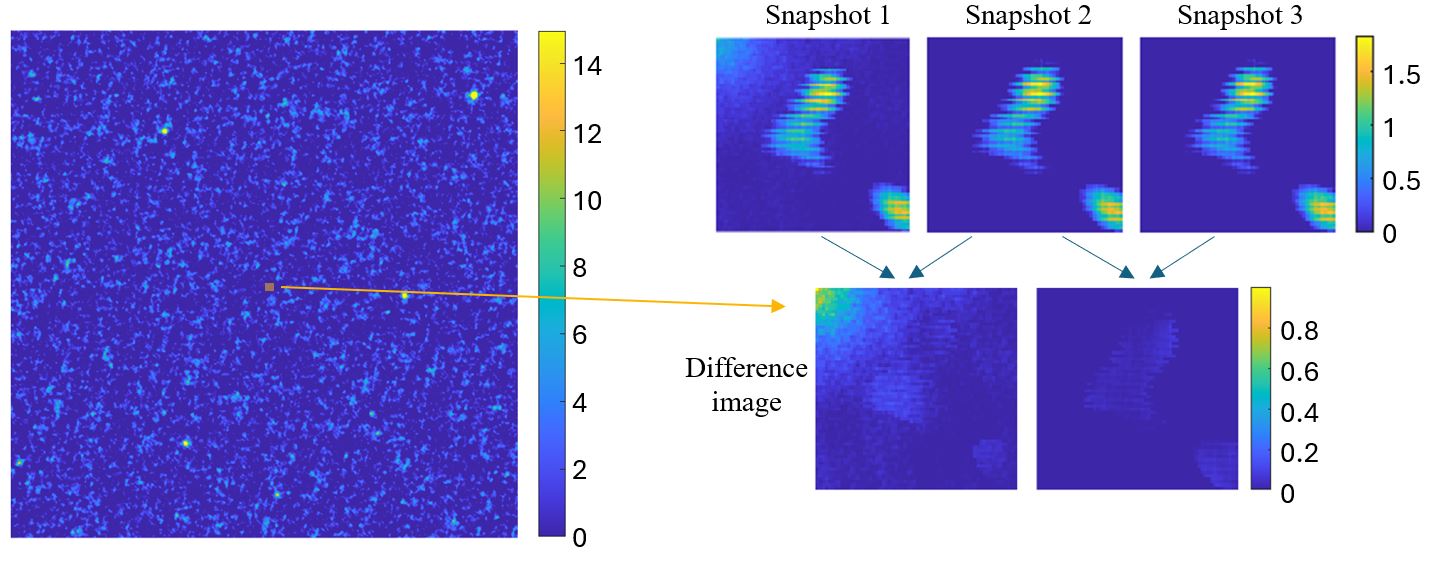}\caption*{(b)}}
\caption{Example images of crops (a) and (b) from the simulated snapshots generated by OSKAR using SKA AA2 telescope layout. In (a), there is no pulsar, while in (b), a pulsar appears in the top-left corner of the tile. Snapshots 1, 2, and 3 are illustrated for both crops. The top row shows the snapshots, and the bottom row shows the difference images between adjacent snapshots.
\label{backgroundplot}}
\end{figure}

As shown in the example in Fig. \ref{backgroundplot}, dynamic steady features can produce intensity differences (from dirty beam convolved on bright steady sources) that surpass the pulsar flux in adjacent difference images. As seen in the figure, crop (a) contains very bright steady features, resulting in significant changes in the difference images, whereas the changes in crop (b) are smaller, even though it contains the true pulsar.  

To assess the significance of these changes quantitatively, we define a ratio $\mathbf{r}$ of intensity variation to the corresponding pixel intensity in the snapshot; the mathematical formulation is presented in Section \ref{sec224}. We apply a threshold of 1 to the ratio $r$ by $r\to\min(r,1)$, ensuring that: 
\begin{itemize}
    \item For pixels with low reference intensity but significant changes, the sensitivity of tLISI remains unaffected;
    \item For pixels with high reference intensity, the sensitivity of tLISI is reduced.
\end{itemize}
This thresholded ratio approach effectively reduces the sensitivity of tLISI to variations in steady features while preserving its sensitivity to pulsar intensity changes. This approach not only reduces the influence of dynamic steady features but also further suppresses residual artefacts that persist after pre-processing (shown in Fig. \ref{addflow}) --- including instrumental noise, radio frequency interference (RFI), and interstellar scattering.

\begin{figure}[!t]
\centering
\begin{minipage}{0.18\textwidth}
\centering
\includegraphics[width=\linewidth]{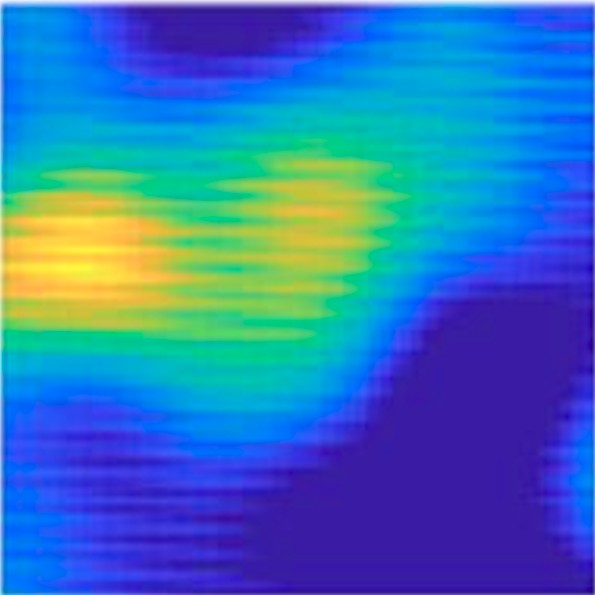}
\end{minipage}
\hfill
\begin{minipage}{0.4\textwidth}
\centering
\includegraphics[width=\linewidth]{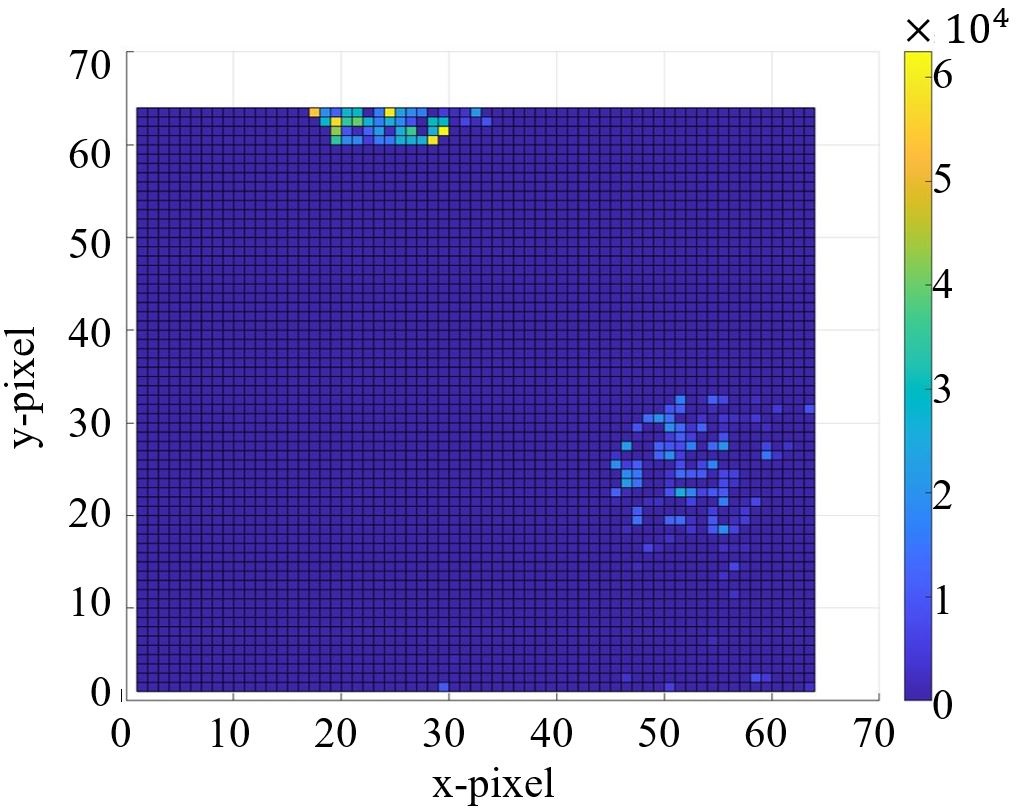}\caption*{(a)}
\end{minipage}
\hfill
\begin{minipage}{0.4\textwidth}
\centering
\includegraphics[width=\linewidth]{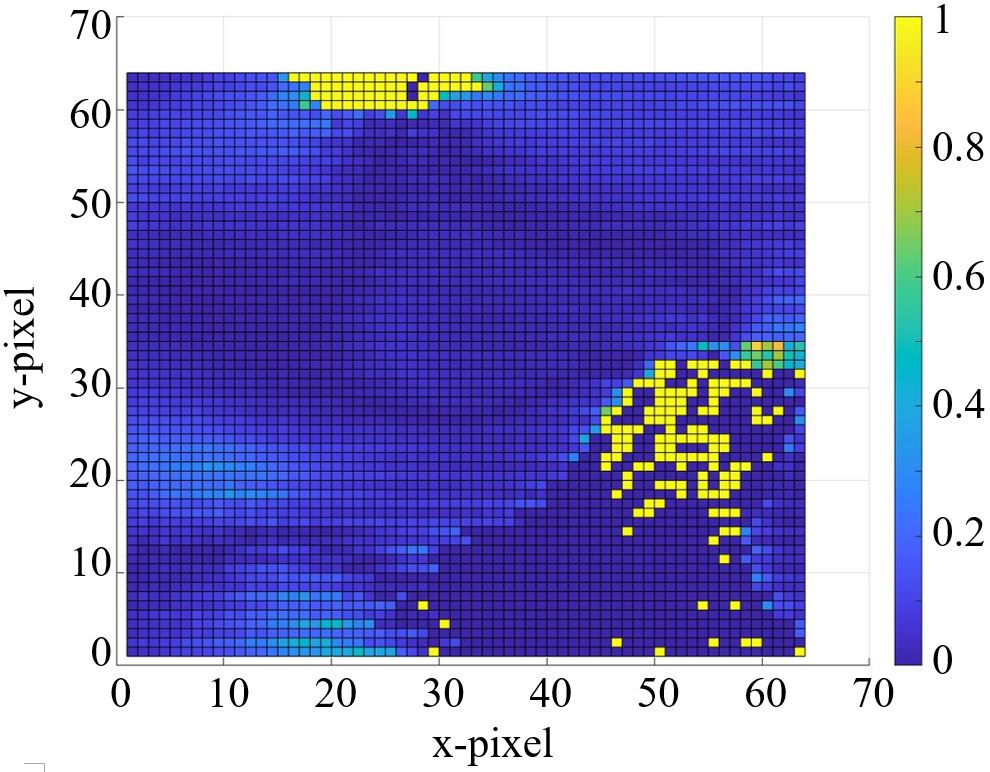}\caption*{(b)}
\end{minipage}
\vspace{0.5em}
\begin{minipage}{0.18\textwidth}
\centering
\includegraphics[width=\linewidth]{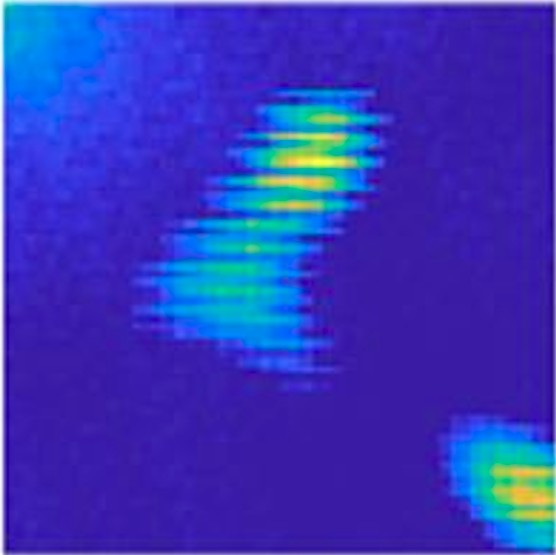}
\end{minipage}
\hfill
\begin{minipage}{0.4\textwidth}
\centering
\includegraphics[width=\linewidth]{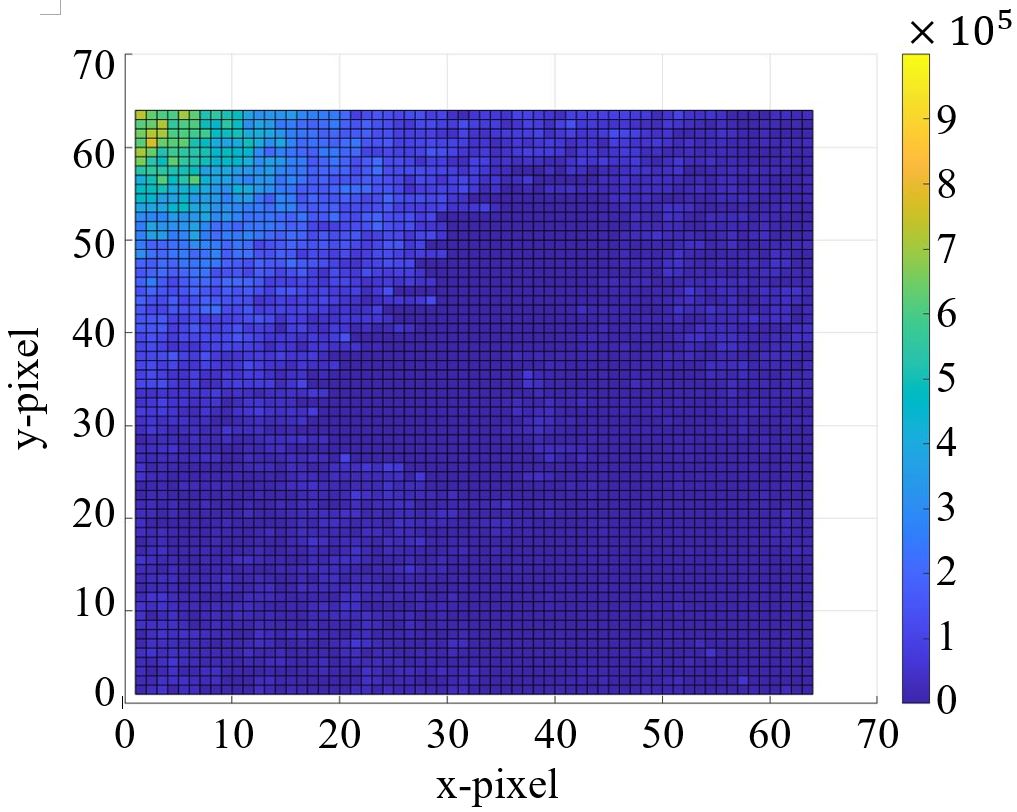}\caption*{(c)}
\end{minipage}
\hfill
\begin{minipage}{0.4\textwidth}
\centering
\includegraphics[width=\linewidth]{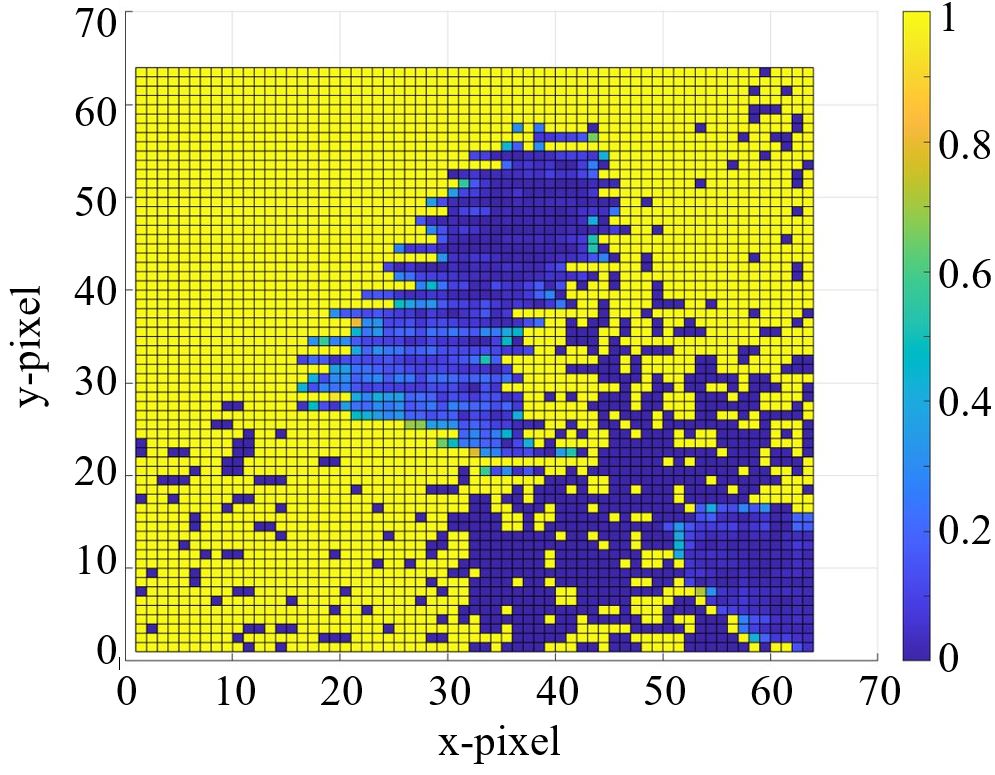}\caption*{(d)}
\end{minipage}
\caption{Effect of the $\mathbf{r}$-term. For clarity, the leftmost column reproduces the two crops of Snapshot 1 shown in Fig. \ref{backgroundplot} (a; top) and (b; bottom). Sub-figures (a) and (b) illustrate the $\mathbf{r}$ values for the crop in Fig. \ref{backgroundplot} (a), before and after applying $r\to\min(r,1)$, respectively, in the absence of a pulsar. On the other hand, sub-figures (c) and (d) demonstrate the corresponding $\mathbf{r}$ values for the crop in Fig. \ref{backgroundplot} (b), which contains a pulsar, again before and after applying $r\to\min(r,1)$. As shown in (d), the thresholded ratio maintains sensitivity to pulsar-induced variations (visible in the top-left corner) while reducing the response to steady features (central part).
\label{separate}}
\end{figure}

To demonstrate the effectiveness of this approach, Fig. \ref{separate} shows that the pulsar will be effectively distinguished from steady features, regardless of whether it is surrounded by brighter or fainter steady features and noise.

\subsubsection{Mathematics}
\label{sec224}

Based on Sections \ref{sec221} to \ref{sec223}, the equation of tLISI is given as
\begin{eqnarray}
    \label{tLISI}
    \mathrm{tLISI}_{t}^{(s)}\left( {\mathbf{D}_{t}^{(s)},\mathbf{D}_{t}^{(s+1)},\mathbf{x}_{t}^{(s+1)}} \right) ~~~~~~~~~~~~~~~~~~~~~~~~~~~~~~~~~~~~~~~~~~~~~~~~~~~~~~~\nonumber\\
    = 1 - \frac{\mathrm{mean}\left( \left| {\mathbf{D}_{t}^{(s)} - \mathbf{D}_{t}^{(s+1)}} \right|\right)\mathrm{max}\left( \left| {\mathbf{D}_{t}^{(s)} - \mathbf{D}_{t}^{(s+1)}} \right|\right)\mathrm{mean}\left( \mathbf{r}_{t}^{(s)}\right) }{\mathrm{max}\left(\mathrm{max}_{s=1}^{L}\mathbf{x}^{(s)}, C \right)\mathrm{max}\left(\mathrm{max}_{s=1}^{L}\mathbf{x}^{(s)}, C \right)},
\end{eqnarray} 
where $t$ is the tile index in an image, $L$ is the total number of snapshots to be analysed, $s = 1, 2, ..., L-2$ is the snapshot index in the time sequential measurement, $\mathbf{x}$ is the snapshot, $\mathbf{D}$ is the difference image between adjacent snapshots which can be expressed by
\begin{equation}
    \mathbf{D}_{t}^{(s)} = \left| \mathbf{x}_{t}^{(s+1)} - \mathbf{x}_{t}^{(s)}\right|,
\end{equation}
$\mathbf{r}$ is a ratio matrix, showing the ratio of intensity difference to its reference intensity level, which can be expressed by
\begin{equation}
\label{equ4}
r_{t,i}^{s} = \left\{ 
    \begin{matrix} 
        {\frac{\left| D_{t,i}^{(s)}-D_{t,i}^{(s+1)}\right|}{\hat{x}_{t,i}^{(s+1)}} ~~, \frac{\left| D_{t,i}^{(s)}-D_{t,i}^{(s+1)}\right|}{\hat{x}_{t,i}^{(s+1)}}<1}\\
        {1~~~~~~~~~, ~\mathrm{otherwise}}\\
    \end{matrix} \right. ,
\end{equation}
where $i$ is the pixel index in the tile and
\begin{equation}
    \hat{x}_{t,i}^{(s+1)} = \left\{ 
    \begin{matrix} 
        {C~~~~~~~~, ~~x_{t,i}^{(s+1)} = 0}\\
        {x_{t,i}^{(s+1)}~, ~\mathrm{otherwise}}\\
    \end{matrix} \right. .
\end{equation}
$C$ is adopted to avoid instability for a very small denominator, which is set to be $10^{-6}$ as it is smaller than the flux (in Jy) of most practical pulsars that have been observed \citep{faintp,faintp2}. In this article, bold variables in mathematical equations denote vectors (arrays or matrices), unless otherwise specified.

To explain the tLISI equation in detail, we divide it into several parts.
\begin{itemize}
\item \textbf{Level of intensity difference}

The level of intensity gap between adjacent difference images is expressed as the normalised mean
\begin{equation*}
\frac{\mathrm{mean}\left( \left| {\mathbf{D}_{t}^{(s)} - \mathbf{D}_{t}^{(s+1)}} \right|\right)}{\mathrm{max}\left(\mathrm{max}_{s=1}^{L}\mathbf{x}^{(s)}, C \right)} = \frac{\Sigma {\left| {\frac{\mathbf{D}_{t}^{(s)}}{\mathrm{max}\left(\mathrm{max}_{s=1}^{L}\mathbf{x}^{(s)}, C \right)} - \frac{\mathbf{D}_{t}^{(s+1)}}{\mathrm{max}\left(\mathrm{max}_{s=1}^{L}\mathbf{x}^{(s)}, C \right)}} \right|}}{\Sigma{1}}.
\end{equation*}

The numerator is divided by the maximum factor $\mathrm{max}\left(\mathrm{max}_{s=1}^{L}\mathbf{x}^{(s)}, C \right)$ to ensure that all images assessed by tLISI have comparable intensity levels. Using the maximum pixel value across all snapshots, rather than only the input pair of difference images, prevents artificially small denominators and resulting undesirably low tLISI values for similar low-intensity images. For real-time processing, this maximum can be estimated from at least three past snapshots. The maximum factor can also be considered as a normalisation factor, ensuring that tLISI remains between 0 and 1.

\item \textbf{Representation of maximum change}

The maximum change between the two adjacent difference images is presented by 
\begin{equation*}
\frac{\mathrm{max}\left( \left| {\mathbf{D}_{t}^{(s)} - \mathbf{D}_{t}^{(s+1)}} \right|\right)}{\mathrm{max}\left(\mathrm{max}_{s=1}^{L}\mathbf{x}^{(s)}, C \right)}
\end{equation*}
to eliminate the influence of the intensity sums within the tiles (to address the case shown by Fig. \ref{affectsum}).

\item \textbf{Emphasis on pulsar change}

To determine the importance of variations, it is necessary to calculate the ratio of the difference $|D_{t,i}^{(s)}-D_{t,i}^{(s+1)}|$ to the intensity level $\hat{x}_{t,i}^{(s+1)}$, expressed as $\mathbf{r}_{t}^{(s)}$. A threshold of 1 is applied to each element $r_{t,i}^{s}$. The ratio can be interpreted as a weighting factor for the importance of changes.

The presence of the weighting factor explains why $\mathbf{x}_{t}^{(s+1)}$ is needed in the input of tLISI. When comparing the $s$-th difference image with the $(s+1)$-th difference image, the $(s+1)$-th snapshot is related to both difference images, thereby serving as the reference image. The $(s+1)$-th snapshot pixel-wise reflects the reference intensity level, regardless of whether both difference images include a pulsar, neither includes a pulsar, or only one includes a pulsar while the other does not (see Fig. \ref{diffmodel}). 

The mean value of the weighting $\mathbf{r}$ is used to measure the average level of importance of variations and to keep the range of tLISI between 0 and 1. The term $\mathbf{r}_{t}^{(s)}$ is crucial in tLISI for distinguishing pulsars from false positives, as shown in Fig. \ref{separate}.
\end{itemize}

\subsection{GPU-Accelerated tLISI}
\label{section22}

As Equation \ref{tLISI} shows, the calculation of each tile is independent in tLISI, allowing for parallel processing to improve computational efficiency.

Compute Unified Device Architecture (CUDA; \citealt{CUDA}) is a parallel computing interface developed by NVIDIA Corporation. It enables users to directly programme NVIDIA GPUs, leveraging the parallel nature of GPUs through CUDA C programming. The CPU \footnote{\url{https://github.com/egbdfX/Intensity-sensitive-IQAs}} and GPU-accelerated \footnote{\url{https://github.com/egbdfX/FastImagingTrigger}} tLISI are open source available on our Github. The tLISI-based FITrig is open-source available on Zenodo \citep{zenodoFIT}.

On the device, the most essential kernel is \texttt{tlisi}, whose pseudo code is shown in Algorithm \ref{tlisikernel}. The target of this kernel is to calculate the tile-wised tLISI in an image with $N \times N$ pixels. According to Equation \ref{tLISI}, $\mathrm{tLISI}$ is a function of $\Delta_{t}^{(s)} = \left| {\mathbf{D}_{t}^{(s)} - \mathbf{D}_{t}^{(s+1)}} \right|$. For the convenience of implementation, Equation \ref{tLISI} is transformed into
\begin{equation}
    \label{tLISI_kernel}
    \mathbf{tLISI}^{(s)} \left( \Delta_{t}^{(s)},\mathbf{x}_{t}^{(s+1)} \right) = 1 - \frac{\Sigma  \Delta_{t}^{(s)}\mathrm{max} \mathbf{\Delta}^{(s)}\Sigma \mathbf{r}_{t}^{(s)} }{{\left(N_T \times N_T\right)}^{2} \times {\mathrm{M}}^{2}},
\end{equation}
where $N_T \times N_T$ is the number of pixels in a tile and $M=\mathrm{max}\left(\mathrm{max}_{s=1}^{L}\mathbf{x}^{(s)}, C \right)$ is the maximum value of pixel intensities. The ``mean'' operation is transformed into a division of the sum, as the ``max'' and ``sum'' operations can be efficiently implemented using reduction operation in CUDA, leveraging shared memory in thread blocks. In this context, a thread block serves as a basic execution unit in parallel computing, where all threads inside the block are able to communicate via shared memory. Thereby, the reduction algorithm efficiently aggregates values from multiple threads by using shared memory to store intermediate results, recursively combining partial outputs in parallel until a final result is obtained. Here, the shared memory is used to compute three values: $\Sigma \Delta_{t}^{(s)}$, $\mathrm{max} \mathbf{\Delta}^{(s)}$, and $\Sigma \mathbf{r}_{t}^{(s)}$.

\begin{algorithm}
\DontPrintSemicolon
\caption{Pseudo code (GPU kernel) for the shared-memory-based tLISI}\label{tlisikernel}
\KwIn{$\Delta^{(s)}$, $\mathbf{x}^{(s+1)}$, $N$, $N_T$, $M$}
\KwOut{$\mathrm{tLISI}$}
\SetKwFunction{FFloorDevice}{floor}
\SetKwFunction{FFmodDevice}{mod}
\SetKwFunction{FCeilDevice}{ceil}
\SetKwFunction{FMax}{max}

\SetKwFunction{FKernel}{tlisi}
\SetKwProg{Fn}{Kernel}{:}{}
\Fn{\FKernel{$\Delta^{(s)}$, $\mathbf{x}^{(s+1)}$, $N$, $N_T$, $M$}}{
    \tcp{Shared memory buffer}
    \_\_shared\_\_  $\mathbf{B_{1,2,3}}$ \tcp*{$\mathbf{B_1}$, $\mathbf{B_2}$, $\mathbf{B_3}$ for $\Sigma \Delta$, $\mathrm{max} \mathbf{\Delta}$, $\Sigma \mathbf{r}$}
    
    \tcp{Indices of the tile}
    $t_x \gets \lfloor \texttt{blockIdx.x}/(N/N_T)\rfloor$, $t_y \gets \mathrm{mod}(\texttt{blockIdx.x}, {N}/{N_T})$
    
    $K \gets \lceil N_T \times N_T / \texttt{blockDim.x} \rceil$
    \BlankLine
    
    \tcp{Segment 1: Assignment}
    \For{$k \gets 1$ \KwTo $K$}{
        \If{$\mathrm{threadIdx.x}+(k-1)\times\mathrm{blockDim.x} < N_T \times N_T$}{
            \If{$k == 1$}{
                ${B}_{1}[\texttt{threadIdx.x}] = 0$
                
                ${B}_{2}[\texttt{threadIdx.x}] = 0$
                
                ${B}_{3}[\texttt{threadIdx.x}] = 0$
            }
            \BlankLine
            \tcp{Indices of pixels}
            $p_x \gets t_x \times N_T + \lfloor (\texttt{threadIdx.x}+(k-1)\times\texttt{blockDim.x})/N_T \rfloor$
            
            $p_y \gets t_y \times N_T + \mathrm{mod}(\texttt{threadIdx.x}+(k-1)\times\texttt{blockDim.x},N_T)$
            \BlankLine
            \tcp{Assignment in shared memory}
            ${B}_{1}[\texttt{threadIdx.x}]$ += $\Delta^{(s)}[p_x \times N + p_y]$\;
            ${B}_{2}[\texttt{threadIdx.x}]$ = \FMax$({B}_{2}[\texttt{threadIdx.x}], \Delta^{(s)}[p_x \times N + p_y])$\;
            \If{$\Delta^{(s)}[p_x \times N + p_y] / \mathbf{x}^{(s+1)}[p_x \times N + p_y] < 1$}{
            ${B}_{3}[\texttt{threadIdx.x}]$ += $\Delta^{(s)}[p_x \times N + p_y] / \mathbf{x}^{(s+1)}[p_x \times N + p_y]$}
            \Else{
            ${B}_{3}[\texttt{threadIdx.x}]$ += 1\;}
        }
        \Else{
            \If{$k == 1$}{
                ${B}_{1}[\texttt{threadIdx.x}] = 0$
                
                ${B}_{2}[\texttt{threadIdx.x}] = 0$
                
                ${B}_{3}[\texttt{threadIdx.x}] = 0$
            }
        }
    }
}
\end{algorithm}

\begin{algorithm}[!t]
\DontPrintSemicolon
\caption{Pseudo code (GPU kernel) for the shared-memory-based tLISI (continue)}\label{tlisikernel2}
\KwIn{$\Delta^{(s)}$, $\mathbf{x}^{(s+1)}$, $N$, $N_T$, $M$}
\KwOut{$\mathrm{tLISI}$}
\SetKwFunction{FFloorDevice}{floor}
\SetKwFunction{FFmodDevice}{mod}
\SetKwFunction{FCeilDevice}{ceil}
\SetKwFunction{FMax}{max}

\SetKwFunction{FKernel}{tlisi}
\SetKwProg{Fn}{Kernel}{:}{}
\Fn{\FKernel{$\Delta^{(s)}$, $\mathbf{x}^{(s+1)}$, $N$, $N_T$, $M$}}{
    
    \tcp{Segment 2: Reduction}
    \For{$\mathrm{d \gets \mathrm{blockDim.x} / 2}$ \KwTo $\mathrm{0}$ \KwBy $\mathrm{d = d / 2}$}{
        \BlankLine
        \tcp{Synchronize threads}
        \_\_syncthreads()
        \BlankLine
        \If{$\mathrm{threadIdx.x} < d$}{
            ${B}_{1}[\texttt{threadIdx.x}]$ += ${B}_{1}[\texttt{threadIdx.x}+d]$\;
            ${B}_{2}[\texttt{threadIdx.x}]$ = \FMax(${B}_{2}[\texttt{threadIdx.x}]$, ${B}_{2}[\texttt{threadIdx.x}+d]$)\;
            ${B}_{3}[\texttt{threadIdx.x}]$ += ${B}_{3}[\texttt{threadIdx.x}+d]$\;
        }
    }
    \BlankLine
    
    \tcp{Segment 3: Output}
    \If{$\mathrm{threadIdx.x} == 0$}{
        tLISI[\texttt{blockIdx.x}] = 1 - $({B}_{1}[0] / N_T / N_T) \times {B}_{2}[0] \times ({B}_{3}[0] / N_T / N_T) / M / M$\;
    }
}
\end{algorithm}

To compute tiles in parallel, each tile is processed within a thread block, with the index of each tile denoted by $(t_x,t_y)$. This kernel consists of three segments.
\begin{itemize}
    \item \textbf{Assignment:} Due to the independence between pixels in tLISI, each thread is responsible for the computations ($\Sigma \Delta_{t}^{(s)}$, $\mathrm{max} \mathbf{\Delta}^{(s)}$, and $\Sigma \mathbf{r}_{t}^{(s)}$) for a set of pixels in the image, indexed by $(p_x,p_y)$: 
\begin{equation}
\left\{\begin{matrix}
    {p_x = t_x \times N_T + \lfloor 
\frac{\texttt{threadIdx.x}+(k-1)\times\texttt{blockDim.x}}{N_T} \rfloor}~~~~~~~~~~~~~~~~~~~~~~~~~~~~\\
    {p_y = t_y \times N_T + \mathrm{mod}(\texttt{threadIdx.x}+(k-1)\times\texttt{blockDim.x},N_T)}
\end{matrix} \right. ,
\end{equation}
where $k = 1, 2, ..., K$ with $K = \lceil N_T \times N_T / \texttt{blockDim.x} \rceil$ such that $\texttt{threadIdx.x}+(k-1)\times\texttt{blockDim.x} < N_T \times N_T$. The corresponding pseudo code is shown in Segment 1 of Algorithm \ref{tlisikernel}.
    \item \textbf{Reduction:} The sums and maximum value of all threads within a block are calculated by the reduction operation, as shown in Segment 2 of Algorithm \ref{tlisikernel}.
    \item \textbf{Output:} In Segment 3 of Algorithm \ref{tlisikernel}, the \texttt{tLISI} value of tile $(t_x,t_y)$ (expressed by \texttt{blockIdx.x}) is calculated and stored.
\end{itemize}

The computational complexity can be analysed by examining each segment. In the assignment stage, each thread independently computes intermediate quantities for a subset of pixels within a tile of size $N_T \times N_T$. Serially, this requires $O(N_T^2)$ operations per tile; however, the independence of pixels allows full parallelisation across threads within a block, reducing the effective per-tile runtime roughly to $O(N_T^2/n_\mathrm{threads})$, where $n_\mathrm{threads}$ is the number of threads per block. Across all tiles in an image of size $N \times N$, the total number of operations in the assignment stage scales as $O(N^2)$. Because of the computations for pixels and tiles are independent, GPU parallelisation allows threads and tiles to be processed concurrently, significantly reducing the wall-closk runtime compared to serial execution.

In the reduction stage, sums and maxima are aggregated across threads using shared-memory reduction. This operation has a logarithmic complexity $O(\mathrm{log}N_T^2)$ per block, which is minor compared to the assignment stage. Finally, the output stage performs a constant-time computation per tile ($O(1)$).

The GPU's highly parallel architecture substantially reduces the effective runtime, making tLISI suitable for rapid processing of large images.

\subsection{Fast Imaging Trigger (FITrig)}
\label{sec22}

FITrig consists of two components: localising potential regions (using tLISI) and pulsar searching (in the image or image-frequency domain). The technique is named ``Trig'' because it operates as a trigger-based detection technique, where tLISI serves as the core mechanism for identifying pulsar signals. FITrig is developed in two complementary versions for pulsar searching --- the fundamental image domain FITrig (Section \ref{section31}) and its enhanced image-frequency domain FITrig (Section \ref{section32}).

\subsubsection{Image Domain Detection}
\label{section31}

\begin{figure}[!t]
    \centering\includegraphics[width=\textwidth]{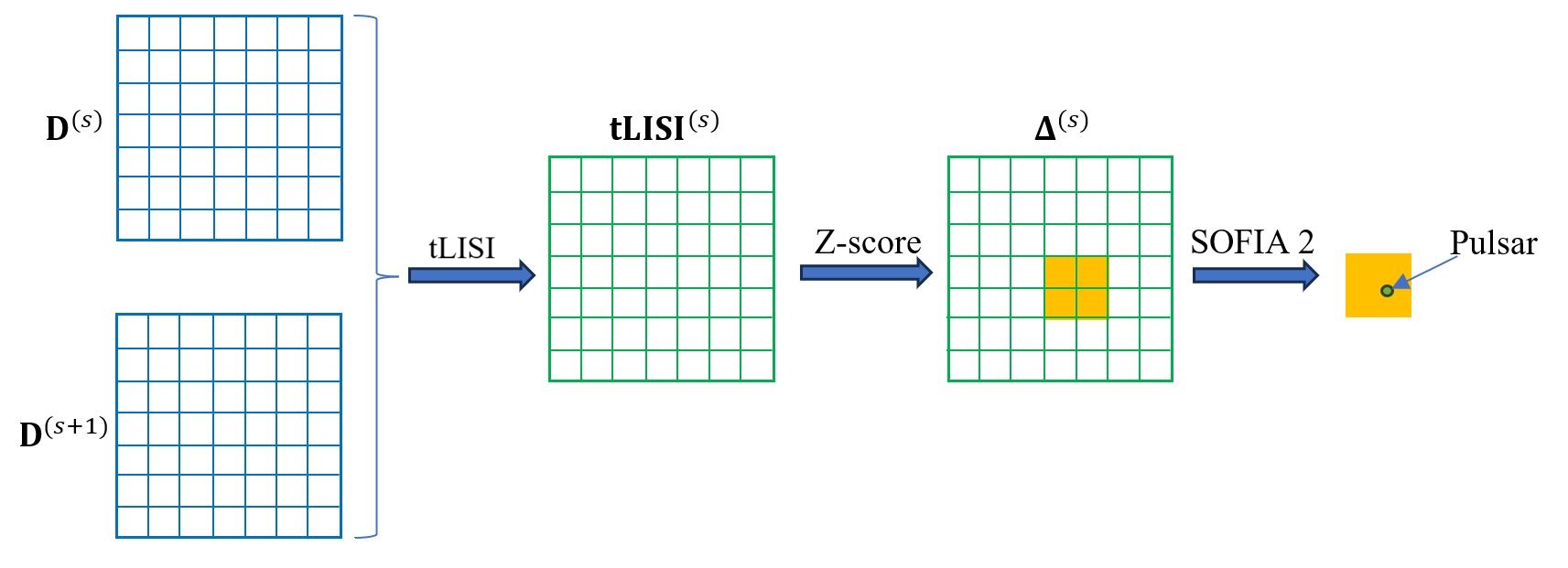}
\caption{Diagram of FITrig: image domain detection approach. Here, the current state-of-the-art source finder, SOFIA 2, is used as an example to demonstrate the functionality of FITrig, though it could be replaced by other source finders.}
\label{imgdetec}
\end{figure}

The image domain detection approach of FITrig is illustrated in Figure \ref{imgdetec}.

The tLISI algorithm outputs a $\frac{N}{N_T} \times \frac{N}{N_T}$ tLISI matrix $\mathbf{tLISI}^{(s)}$ that indicates the similarity between the difference images $\mathbf{D}^{(s)}$ and $\mathbf{D}^{(s+1)}$. Each element of the tLISI matrix denotes the tLISI value of the corresponding tile in the difference images. Subsequently, to statistically identify candidate tiles likely to contain pulsar(s), the z-score is calculated for the tLISI matrix, as shown in
\begin{equation}
z_{t}^{(s)} = \frac{(1-\mathrm{tLISI}_{t}^{(s)}) - (1-\mu^{(s)})}{\sigma^{(s)}} =\frac{\mu^{(s)}-\mathrm{tLISI}_{t}^{(s)}}{\sigma^{(s)}},
\end{equation}
where $t$ is the tile (element) index in the tLISI matrix and z-score matrix, $z_{t}$ is the z-score for the $t$-th tile, $\mathrm{tLISI}_{t}$ is the tLISI value for the $t$-th tile, and $\mu$ and $\sigma$ represent the mean and standard deviation of the tLISI matrix, respectively. Note that the complement of tLISI is used in the calculation of the z-score because the greater the difference between the two images, the closer the tLISI value is to 0. Therefore, the similarity of the two images is represented by the value of tLISI, while the difference is represented by the value of the complement of tLISI. 

The tiles of the image with z-scores higher than a threshold $\tau$ (discussed in Section \ref{section4}) will be selected as candidate tiles. Typically, a pulsar is often included in more than one tile, especially when the tile size is very small. In such cases, the tLISI step will forward all relevant tiles to the source finder, e.g., SOFIA 2, for accurate localisation of the pulsar. It will output the positions of pulsars in Right Ascension and Declination (RA/Dec), along with their fluxes. The set of indices for the selected tiles is expressed as 
\begin{equation}
    \left\{ t | z_t^{(s)} > \tau \right\}.
\end{equation}

\subsubsection{Image-Frequency Domain Detection}
\label{section32}

As analysed in Section \ref{section21}, ULP pulsars exhibit sparse temporal distributions. For detecting these pulsars in datasets containing unknown sources, the image-frequency domain FITrig proposed in this section provides a more robust localisation solution, building upon the fundamental image domain FITrig (Section \ref{section31}).

\begin{figure}[!t]
    \centering
    \includegraphics[width=\textwidth]{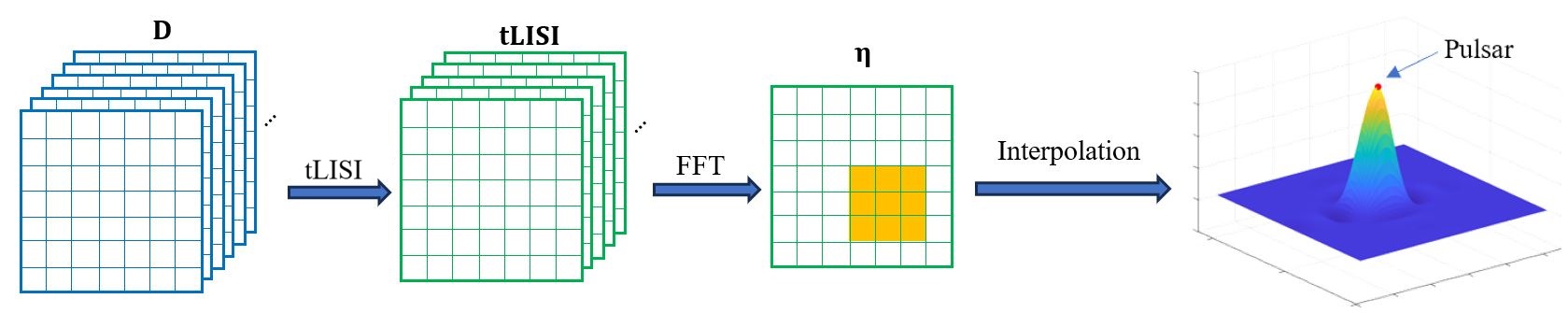}
\caption{Diagram of FITrig: image-frequency domain detection approach.}
\label{freqdetect}
\end{figure}

Based on the periodicity of ULP pulsars, Fig. \ref{freqdetect} shows the image-frequency domain detection approach of FITrig.

In this approach, the tLISI algorithm is applied to time-sequential difference images $\mathbf{D}$ to obtain a $\frac{N}{N_T} \times \frac{N}{N_T} \times (L-2)$ tLISI cube, constructed from tLISI values for tiles in each pair of adjacent difference images, with the third dimension representing time. 

FITrig executes a fast Fourier transform (FFT) on the sequential tLISI values corresponding to each tile over the time axis. This process results in a 3-dimensional (3-D) cube, where the first two axes represent the positions of tiles and the third dimension represents frequency. Subsequently, a spectrum matrix $\mathbf{\eta}$ is constructed, with each element $\eta_t$ representing the informative magnitude of the frequency spectrum along the third axis for the corresponding tile $t$, as shown in
\begin{equation}
    \eta_t = h\left( \left|\mathcal{F}\left[\mathrm{tLISI}_t(s)-\mathrm{mean}\left(\mathrm{tLISI}_t(s)\right)\right]\right|\right), 
\end{equation}
where $\mathcal{F}[\bullet(s)]$ indicates the Fourier transform operation over the $s$-axis and $h$ indicates a spectral refinement function, such as harmonic summing \citep{harmonic1}, to form informative spectrum. The mean of the $\mathrm{tLISI}_t$ sequence, represented by the Direct Current (DC) component from the spectrum, is removed. 

The values in the spectrum matrix $\mathbf{\eta}$ act as contributions of the pulsar(s) to different tiles. Typically, a pulsar primarily contributes to either one tile, a cluster of two tiles (when the pulsar is on the boundaries of tiles), or a cluster of four tiles (when the pulsar is at the corners of tiles). All potential tiles that may contain pulsar(s) are selected by applying a threshold $\tau$ (discussed in Section \ref{section4}) on the z-score matrix of the spectrum matrix, as expressed by 
\begin{equation}
    z_{t} = \frac{\eta_t - \mathrm{mean}(\mathbf{\eta})}{\mathrm{std}(\mathbf{\eta})}.
\end{equation}
The set of indices for the selected tiles is given by
\begin{equation}
    \left\{ t | z_t > \tau \right\}.
\end{equation}

To reconstruct the pulsar position based on its contributions across different tiles, interpolation methods are activated for precise pulsar localisation in the selected regions. Unlike pixel-level methods such as 2-D Gaussian fitting that operate on snapshots, our tile-based approach using bilinear, spline, bicubic, or Makima cubic interpolation offers computational advantages while maintaining precision. This method is effective for real-time automation of localising ULP pulsars in large-area surveys.

The image-frequency domain FITrig takes into account the time-sequential tLISI, which further reduces the influence of noise.

\section{Performance}
\label{section4}

We evaluate FITrig in two main settings. The first involves testing the real-world measurement of PSR J0901-4046 to examine its performance under practical conditions and realistic data variability. The second comprises a series of experiments on simulated datasets, providing controlled settings to evaluate FITrig's robustness across diverse observational conditions.

\subsection*{\textbf{Real Dataset Evaluation}}

\begin{figure}[!t]
\centering
\begin{minipage}{0.4\textwidth}
\centering
\includegraphics[width=\linewidth]{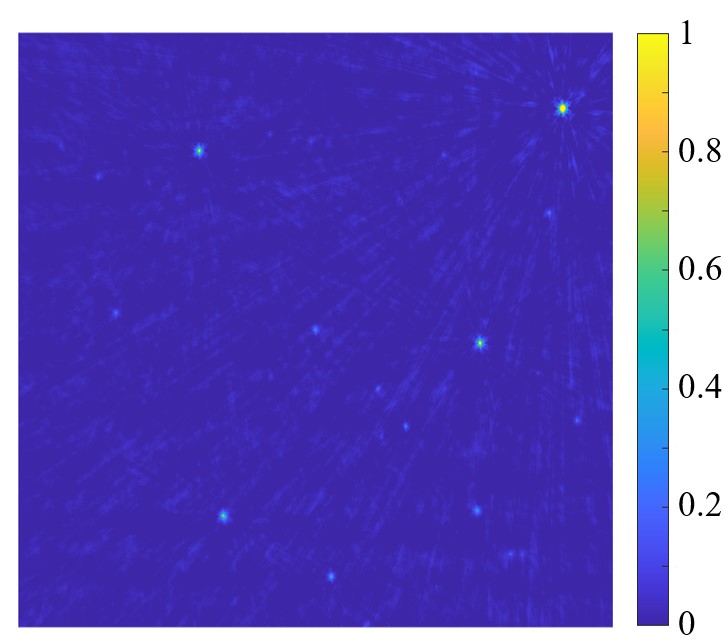}\caption*{(a)}
\end{minipage}
\hfill
\begin{minipage}{0.5\textwidth}
\centering
\includegraphics[width=\linewidth]{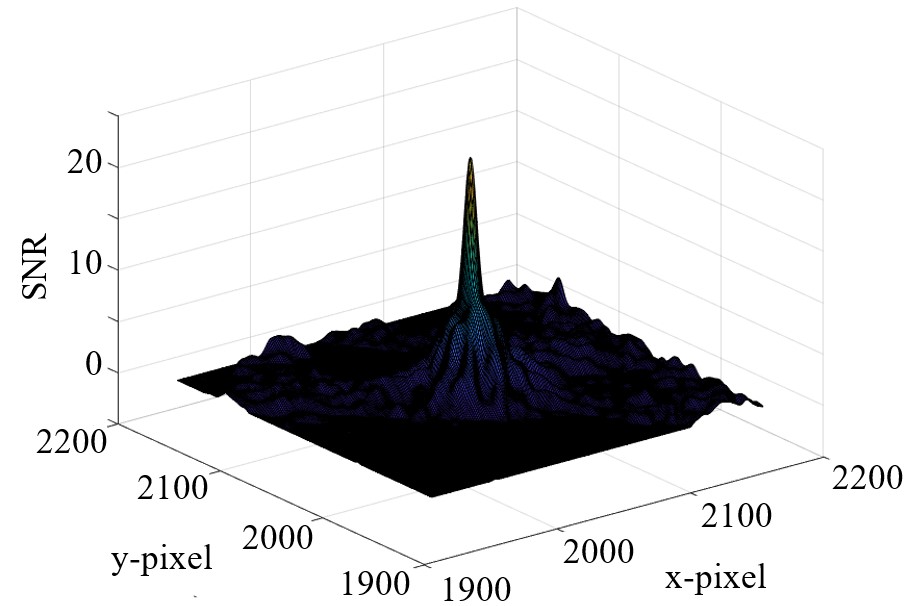}\caption*{(b)}
\end{minipage}
\caption{FOV centred on PSR J0901-4046, observed with MeerKAT, with an image size of $4096 \times 4096$ pixels. Intensities are normalised for illustration purposes. (a) Dirty snapshot in which the pulsar exhibits its highest intensity within the dataset. (b) SNR plot corresponding to (a), showing only the regions around the pulsar. \label{realdirty}}
\end{figure}

We evaluate FITrig on a real dataset of PSR J0901-4046, a periodic pulsating neutron star with a period of 76 seconds, observed by MeerKAT in the L band \citep{realpul2}. The Measurement Set contains 1500 time samples of visibilities, with a sampling period of 2 seconds. Figure \ref{realdirty} shows a 1.2-degree Field of View (FOV) centred on the pulsar in sub-figure (a), along with the corresponding SNR plot in sub-figure (b), where the pulsar reaches an SNR of 22.7.

\begin{figure}[!t]
    \centering
    \includegraphics[width=0.5\textwidth]{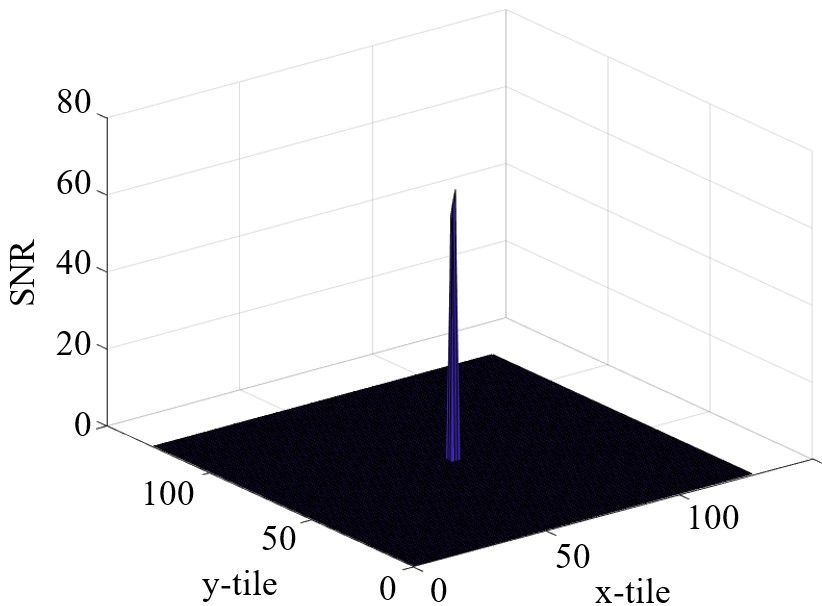}
\caption{SNR plot of spectrum matrix $\mathbf{\eta}$ obtained by applying FITrig ($N_T=32$) to the real dataset of PSR J0901-4046.}
\label{etareal}
\end{figure}

To search for the ULP pulsar in the dataset, we use FITrig, whose processing steps are illustrated in Fig. \ref{freqdetect}, with harmonic summing applied to the spectrum. Figure \ref{etareal} shows the SNR plot of the spectrum matrix $\mathbf{\eta}$, in which the pulsar appears as a distinct peak with an SNR of 70.3, higher than the SNR in any snapshot.

We apply spline interpolation to the spectrum matrix to strike a balance between precision and computational efficiency \citep{xiaotongthesis}. The interpolation ensures that the peak position, corresponding to the pulsar position, aligns with its actual coordinates at RA 135.37$^\circ$, Dec -40.77$^\circ$, yielding arcsecond-level localisation accuracy. After detecting the ULP pulsar from the dataset with FITrig, higher localisation precision can also be achieved by activating a source finder, such as SOFIA 2, on a tile containing the pulsar. The resulting localisation accuracy then depends on the precision of the source finder.

\begin{figure}[!t]
    \centering
    \includegraphics[width=\textwidth]{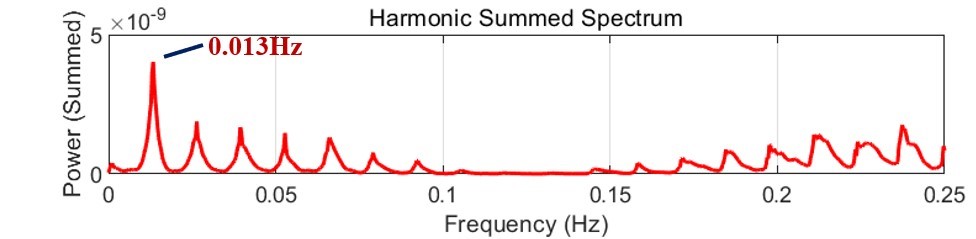}
\caption{Harmonic summed spectrum corresponding to the time-sequential images of the pulsar tile. The fundamental frequency in the spectrum is 0.013 Hz, corresponding to the spin frequency of PSR J0901-4046.}
\label{pulsarfft}
\end{figure}

Focusing on the pulsar tile, the harmonic summed spectrum (Fig. \ref{pulsarfft}) clearly reveals a frequency of 0.013 Hz, corresponding to a 76-second period. This demonstrates that FITrig not only localises the pulsar with high precision but also recovers its accurate spin frequency, further confirming the effectiveness of FITrig.

In addition to its accuracy, FITrig offers a substantial advantage in computational efficiency. In the experiments, FITrig is executed on the NVIDIA H100 \footnote{\url{https://resources.nvidia.com/en-us-tensor-core}}. With high levels of parallelisation and GPU acceleration, FITrig can complete pulsar detection and localisation in 6.34 seconds on this 1500-snapshot dataset, covering the process from image loading to pulsar position output. This performance is superior to that of existing pulsar detection techniques and can be regarded as real-time. For details of the computational efficiency of FITrig, a more comprehensive discussion is provided in Section \ref{CompEffi}. This includes consideration of very large-scale images (e.g., 50K $\times$ 50K pixels), as next-generation telescopes can produce data at this scale, where existing methods are computationally challenging for pulsar searches on such large images.

As shown by the real-data evaluation, FITrig demonstrates accurate localisation of the ULP pulsar and precise frequency detection at very high speed. In fact, beyond computational efficiency, it also achieves a higher sensitivity (Section \ref{413}) and a lower false positive rate (Section \ref{sec41}) than existing methods, which will be further evaluated on simulated datasets under a variety of conditions in the following section.

\subsection*{\textbf{Simulation-Based Evaluation}}

\subsection{Parameters and Experimental Settings}
\label{22p}

We evaluate FITrig using simulated datasets across various settings and parameter configurations. To avoid bias from different telescope layouts, we simulate images for the same SBD using telescope configurations of MeerKAT and SKA AA2. The sampling rate is tested across a range from 0.016 seconds (or smaller) to 16 seconds, where the lower bound exceeds the capabilities of current telescopes as our design targets next-generation telescopes or beyond, while the upper bound exceeds the sampling rates achievable by current techniques, demonstrating the practicality of our method for existing telescopes.

To evaluate FITrig performance under varying conditions, the three-snapshot unit (as shown in Fig. \ref{tiles}) is created following three key principles. Firstly, images are based on the real Measurement Set to represent practical scenarios. Specifically, we image and deconvolve snapshots from the real Measurement Set, then use the resulting restored image as a sky model in OSKAR to produce simulated Measurement Sets using MeerKAT and SKA AA2 telescope layouts, separately. Secondly, images are designed to simulate the behaviour of ULP pulsars during observations. To be more precise, in the three-snapshot unit, the pulsar is only visible in one of the three snapshots. W-Stacking Clean (WSClean; \citealt{WS1}) is used to reconstruct dirty snapshots from Measurement Sets. Thirdly, several sets of images are produced with steady features at different intensity levels. More concretely, in the simulation, the relative brightness between steady features and the pulsar is controlled by $\Gamma = 0.05, 1, 20, 40, 60, 80, 100$, defined as the ratio of the brightest steady features to the pulsar flux.

Three important parameters (threshold $\tau$, tile size $N_T \times N_T$, length of time-sequential snapshots $L$) are explained below. To begin with, the parameter $\tau$ serves as the threshold for z-score values, filtering significant cases by indicating how many standard deviations ($\sigma$) the tLISI of pulsar deviates from the average level ($\mu$). We investigate sigma levels ranging from 3 to 6, with 5$\sigma$ as the reference level as it is commonly used in transient detection studies \citep{m81,sourcefinder1}.

With regard to the second parameter, tile size, we use the number of pixels to represent effective tile size $N_T$, rather than the degrees of FOV covered by a tile, to avoid the influence of different telescope resolutions. Table \ref{table005} illustrates the effective tile sizes for different combinations of threshold $\tau$, telescope arrays, and relative brightness $\Gamma$. The ``-'' entry indicates that, for the given parameters, no valid effective tile size can be determined. Here, the phrase ``effective tile sizes'' refers to tile sizes for which the pulsar tile is successfully selected while keeping the proportion of selected pixels below a level of $\theta=0.5\%$ \citep{xiaotongthesis}. This level $\theta$ controls the trade-off between false positives and detection sensitivity --- lower $\theta$ reduces false positives but also decreases sensitivity. For example, when $\theta=0.1\%$ and steady sources are very bright ($\Gamma>20$), the pulsar becomes undetectable in our experiments.
\begin{table}[t]
\centering
\begin{tabular}{ccccccccc}
\hline
\multicolumn{2}{c}{Effective $N_T$} & \multicolumn{7}{c}{\textbf{$\Gamma$}}\\
\textbf{Telescope} & \textbf{$\tau$} & \textbf{0.05} & \textbf{1} & \textbf{20} & \textbf{40} & \textbf{60} & \textbf{80} & \textbf{100}\\
\hline
\textbf{MeerKAT} & \textbf{3} & 1-128 & 1-128 & 1-64 & - & - & - & - \\
  & \textbf{4} & 1-128 & 1-128 & 1-64 & 32-64 & - & - & - \\
  & \textbf{5} & 1-128 & 1-128 & 1-64 & 8-64 & - & - & - \\
  & \textbf{6} & 1-128 & 1-128 & 1-256 & 1-128 & 1-64 & 1-64 & 1-64\\
\textbf{SKA AA2} & \textbf{3} & 1-128 & 1-64 & 2-64 & - & - & - & - \\
  & \textbf{4} & 1-128 & 1-128 & 1-128 & 16-32 & - & - & - \\
  & \textbf{5} & 1-128 & 1-128 & 1-128 & 1-64 & 32 & - & - \\
  & \textbf{6} & 1-128 & 1-128 & 1-256 & 1-128 & 1-64 & 1-64 & 1-32 \\
\hline
\end{tabular}
\caption{Effective tile sizes under different choices of circumstances (telescope layouts, significance threshold $\tau$, pulsar faintness $\Gamma$), with the number of potential false positives to be less than $\theta=0.5\%$ number of pixels of the whole image.
\label{table005}}
\end{table}

In the results, the pulsar detected by the FITrig can be very faint, with its flux reaching between 1/40 and 1/60 of that of the surrounding steady features in the images using MeerKAT and SKA AA2, respectively, for $\tau = 5$. A very small tile size (e.g., per-pixel detection) amplifies the impact of SNR fluctuations. Comprehensively, the best tile size for pulsar localisation is 32 $\times$ 32 pixels. While we adopt this size, FITrig supports customisable tile sizes to accommodate diverse user needs.

Regarding the third parameter $L$, longer observation durations (larger $L$ at constant sampling period $T_s$) enhance signal detection through greater energy accumulation and reduced noise, but require processing more snapshots, creating a real-time performance trade-off.

In the experiments, we simulate three pulsars under different sampling conditions: one undersampled, one nearly critically sampled, one oversampled, according to the Nyquist criterion. In the experiment, various lengths of snapshots are performed, including $L$ = 258, 514, 1026, and 2050, ensuring that the number of sequential tLISI matrices over time ($L-2$) is a power of 2 for computational purposes. The pulsar profile is simulated as a truncated Gaussian \citep{Pulsar} to avoid power leakage between pulsar periods. The duty cycle of the pulsar is set to 10\% of the pulsar period. Within each sampling period, the pulsar intensities are accumulated and integrated to obtain flux.

Please note that in this article we demonstrate FITrig on typical on-and-off pulsars; a more comprehensive evaluation of its generalisation, including cases with multiple transients in the same FOV and pulsars exhibiting gradual variations in brightness over time, will be presented in an upcoming paper on the Fast Imaging Pipeline in this series.

\subsection{Computational Efficiency}
\label{CompEffi}

In FITrig, each tile is computed in parallel, as explained in Section \ref{section22}. In image domain FITrig, SOFIA 2 only needs to process a few tiles rather than the whole image, which effectively speeds up the computing process. The enhancement is more evident when the image is larger. To evaluate computational performance, we input images filled with random Gaussian noise. This approach ensures consistency, because the computation time of FITrig depends solely on pixel intensities and remains unaffected by image content. In contrast, SOFIA 2's runtime varies depending on the number and complexity of sources it detects, since it performs feature extraction. Using these inputs, we eliminate content-related variables and focus solely on comparing the underlying computational efficiency of both methods.

In the experiments, images with sizes ranging from 512 $\times$ 512 pixels to 50K $\times$ 50K pixels are inputted separately. For each image size, 10 tests are run and the median computing time among the 10 tests is calculated. Three consecutive snapshots are inputted for each run. Since the tiles selected by the tLISI algorithm will be forwarded for further localisation of the pulsars, SOFIA 2 will run on the selected tiles.

\begin{figure}[!t]
\centering
\begin{minipage}{0.45\textwidth}
\centering
\includegraphics[width=\linewidth]{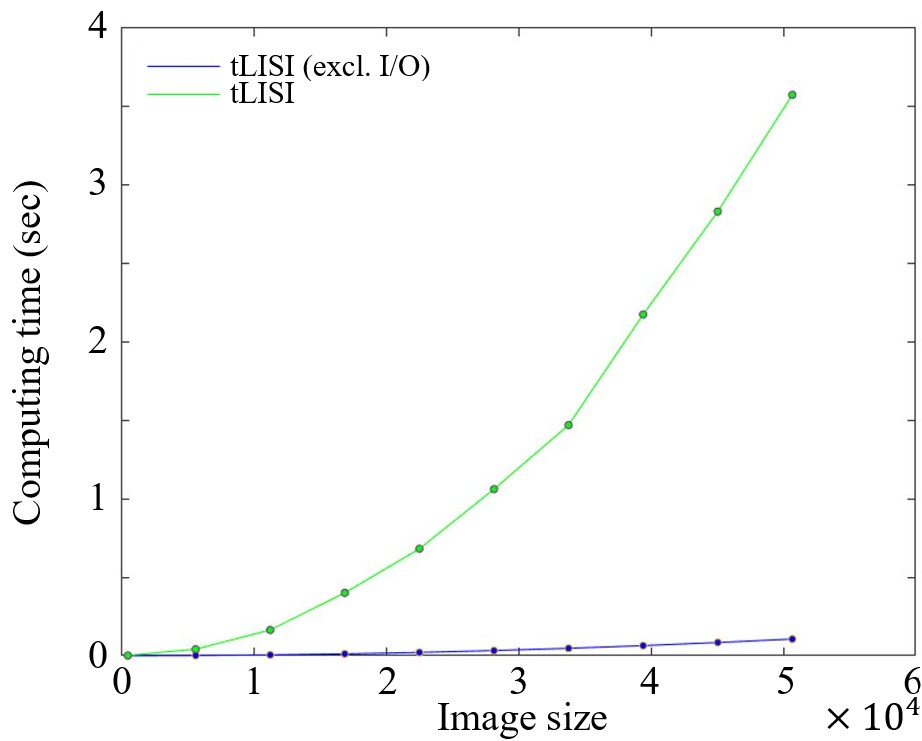}\caption*{(a)}
\end{minipage}
\hfill
\begin{minipage}{0.45\textwidth}
\centering
\includegraphics[width=\linewidth]{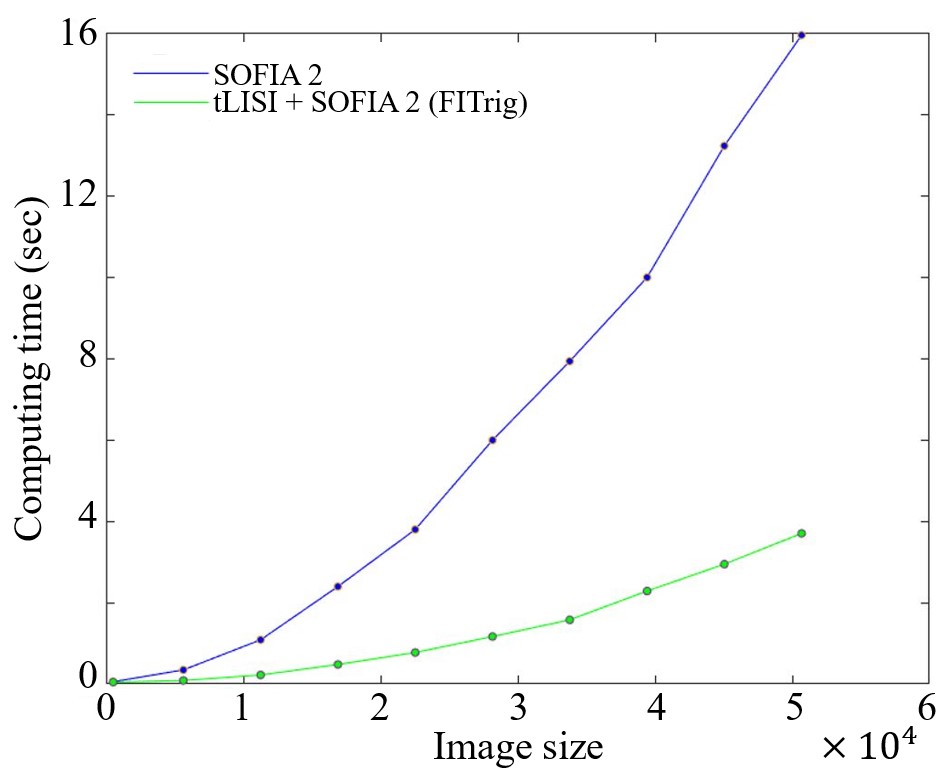}\caption*{(b)}
\end{minipage}
\caption{Computation time, where (a) compares the computation time excluding (blue lines) and including (green lines) I/O time between host \& device and memory allocation time, and (b) compares the computation time for FITrig (green lines) and using SOFIA 2 alone (blue lines). The image size is represented by $N$. \label{timeimage}}
\end{figure}

The results of the computation time are shown in Fig. \ref{timeimage}. The computation time of the tLISI algorithm, excluding I/O time between host and device as well as memory allocation time, is also shown in the results (as ``excl. I/O''). This is because when the tLISI is applied in the middle of a FI pipeline, the data is already available on the device, so there is no need to carry out memory allocation and data transfer between the host and the device.

According to the results, when the image size increases, our FITrig speeds up processing, which can be 4.3 times faster than solely applying SOFIA 2 when the image size reaches 50K $\times$ 50K pixels.

In image-frequency domain FITrig, a set of time-sequential snapshots needs to be loaded onto the device. Rather than loading all snapshots at once, a better way to compute is to cover the data transfer time on Peripheral Component Interconnect Express (PCIe) as much as possible by the computing process of tLISI. CUDA kernels running on one group of inputs (three consecutive snapshots) are called ``trigger kernel''. The output of each trigger kernel is a tLISI matrix.

\begin{figure}[!t]
    \centering
    \includegraphics[width=0.65\textwidth]{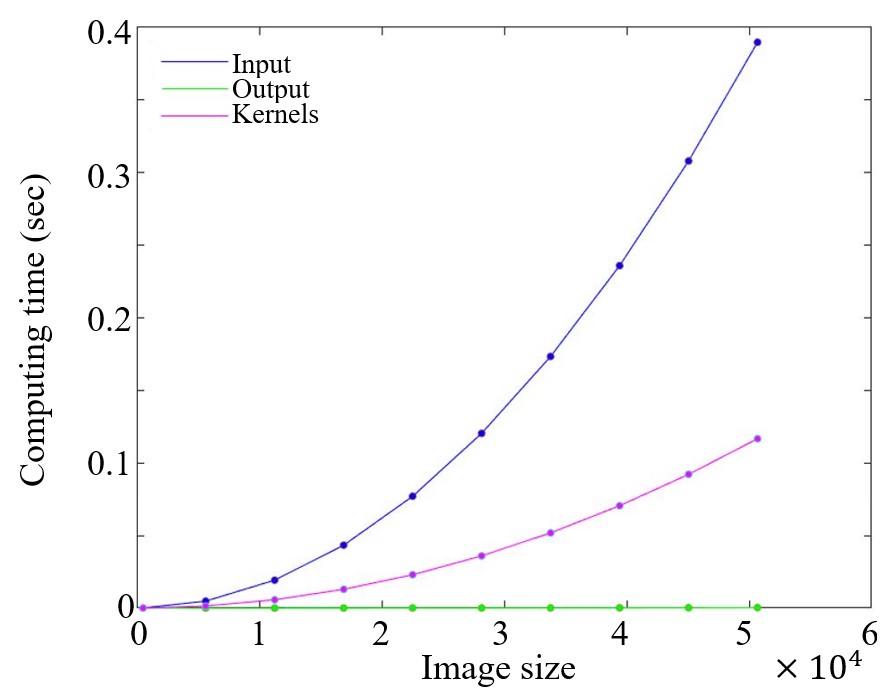}
\caption{Time of data transfer (input and output) with pinned memory vs. time of the trigger kernel against image size ($N$).}
\label{pciemeasure}
\end{figure}

To design the computing process, we measure the time of transferring a snapshot (input) and a tLISI matrix (output) over PCIe Gen 4.0 using pinned memory. The time of data transfer and the time of trigger kernel for different sizes of images are shown in Fig. \ref{pciemeasure}. Assuming the same utilisation, switching from PCIe 4.0 to PCIe 5.0 will double the speed due to the increased bandwidth.  

\begin{figure}[!t]
    \centering
    \includegraphics[width=\textwidth]{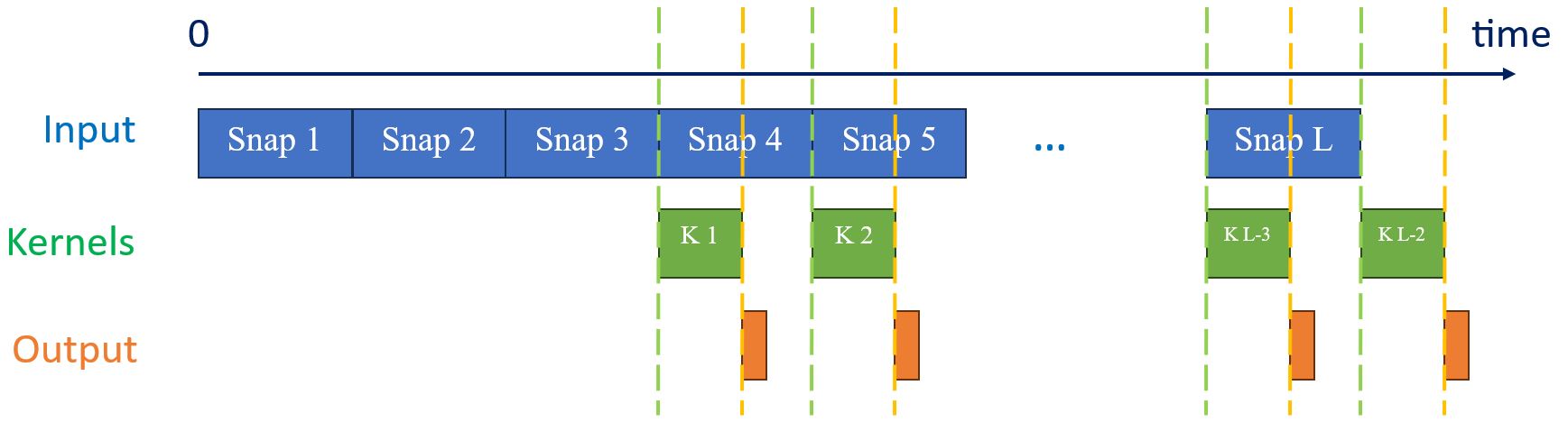}
\caption{Parallel timeline of the trigger: ``Snap $s$'' ($s = 1, 2, ..., L$) indicates the $s$-th snapshot, ``K $n$'' ($n = 1, 2, ..., L-2$) indicates the $n$-th trigger kernel process, and the orange bar (too small for a label) indicates the $n$-th ($n = 1, 2, ..., L-2$) output tLISI matrix. ``Input'' indicates the process of transferring a snapshot from the host to the device, ``Kernels'' indicates the execution of trigger kernels on the device, and ``Output'' indicates the process of transferring a tLISI matrix from the device to the host.}
\label{pciediagram}
\end{figure}

As shown in Fig. \ref{pciemeasure}, the time required to output a tLISI matrix is very short, and can be ignored compared to the kernel processing time and the input transfer time. To improve efficiency, the design of the parallel timeline is shown in Fig. \ref{pciediagram}. Specifically, three consecutive snapshots are needed for each tLISI computation. At the beginning, the first three snapshots are loaded onto the device. Then, while the $n$-th ($n = 1, 2, ..., L-3$) tLISI is being computed, the $s=n+3$ th snapshot is loaded. The $n+1$ th tLISI computation begins once the $s = n+3$ th snapshot has finished loading. The $n=L-2$ th tLISI will be the last, during which no further data is loaded. The $n$-th ($n = 1, 2, ..., L-2$) tLISI result is transferred once the $n$-th tLISI process is done.

\subsection{Sensitivity}
\label{413}

Current telescopes like MeerKAT can achieve sampling periods of 2 or 8 seconds, while next-generation telescopes such as the SKA will reach 1-second sampling period \citep{ska1low}. Future telescopes may further reduce this duration. Detection sensitivity is related to the sampling period $T_s$ as it affects signal integration within a snapshot. Sections \ref{sec431} and \ref{sec432} analyse detection strategies and sensitivity for long and short sampling periods --- where ``long'' and ``short'' are relative descriptions. The goal is to determine FITrig's maximum tolerable sampling period and to explore mitigation strategies for scenarios where future telescopes employ extremely short sampling periods. Section \ref{sec42} presents the detection sensitivity of FITrig across varying sampling periods corresponding to different Nyquist sampling conditions.

\subsubsection{Detection Sensitivity Analysis for Long Sampling Periods}
\label{sec431}

\begin{figure}[!t]
\centering
\begin{minipage}{0.45\textwidth}
\centering
\includegraphics[width=\linewidth]{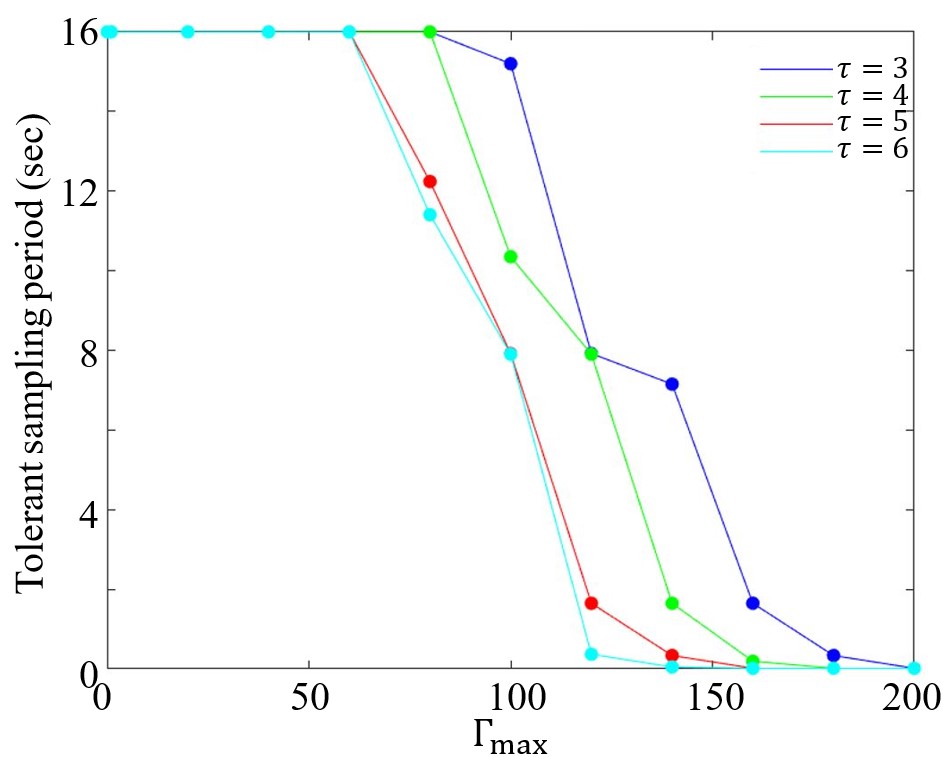}\caption*{(a)}
\end{minipage}
\hfill
\begin{minipage}{0.45\textwidth}
\centering
\includegraphics[width=\linewidth]{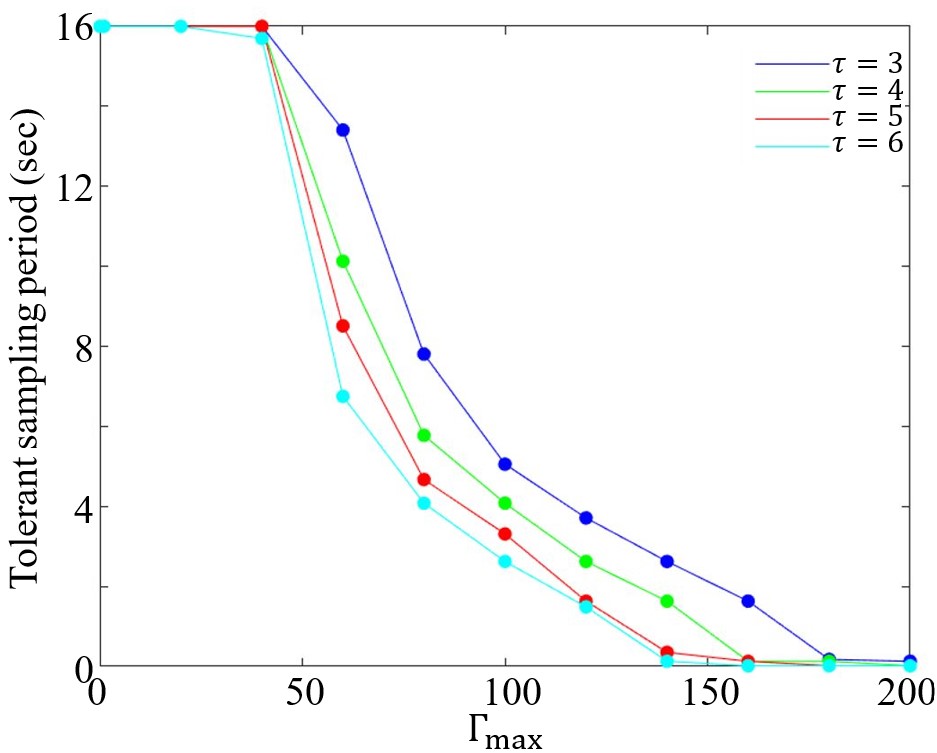}\caption*{(b)}
\end{minipage}
\caption{Allowable sampling period under various settings of $\Gamma$ and $\tau$ using telescope array (a) MeerKAT and (b) SKA AA2.
\label{timediff}}
\end{figure}

Larger $T_s$ values lead to increased variations in steady features of consecutive snapshots. Therefore, it is crucial to determine the maximum allowable $T_s$ when applying the FITrig. As shown in Fig. \ref{timediff}, experiments reveal the maximum allowable $T_s$ for various $\Gamma$ values using configurations of MeerKAT and SKA AA2. The allowable sampling period is obtained by comparing the z-score of pulsar tile to a threshold $\tau$. If the z-score surpasses $\tau$ (i.e., the pulsar is detected), the corresponding sampling period is considered ``allowable''; otherwise, it is not. In the experiments, the maximum sampling period evaluated is 16 seconds, which is sufficient to show the cases for pulsar detection.

The findings indicate that the appropriate sampling period for applying FITrig in detection is influenced by both the telescope configuration and the desired sensitivity. The allowable sampling period is constrained by the intensity of steady features in the same SBD. For instance, detecting a pulsar 100 times fainter than the brightest surrounding steady features, and using a threshold of 6, the sampling period must not exceed 7.9 seconds for MeerKAT and 2.6 seconds for SKA AA2.

\subsubsection{Sensitivity Limit and Mitigation Approach for Short Sampling Periods}
\label{sec432}

An extremely short $T_s$ --- especially with advanced telescopes that offer ultra-high sampling rates --- results in consecutive snapshots being almost identical, whether they capture transient or steady features. In such scenarios, applying FITrig to near-blank difference images is impractical.

\begin{figure}[!t]
    \centering
    \includegraphics[width=\textwidth]{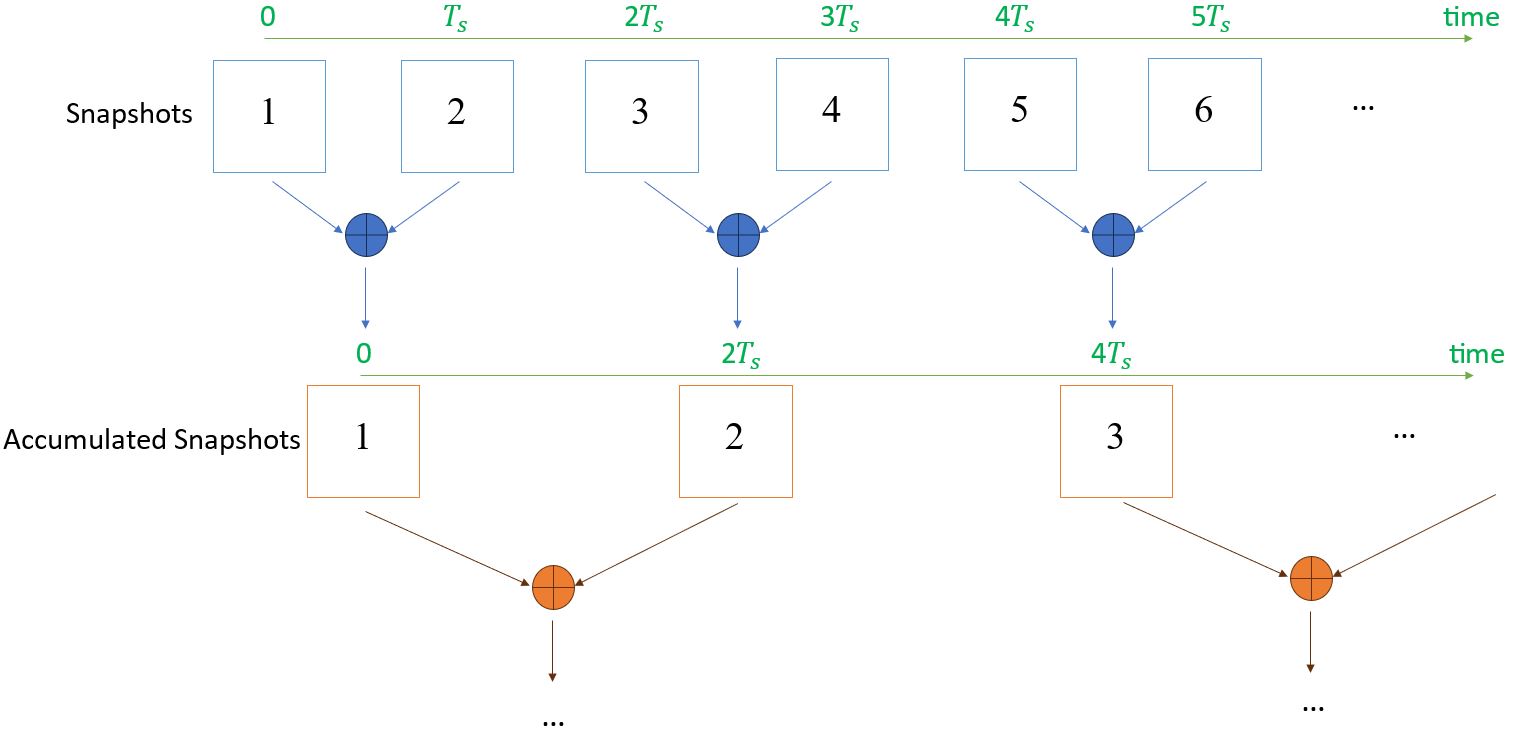}
\caption{Snapshots accumulation, where the symbol of ``$\oplus$'' means adding the two snapshots together and outputting their sum.
\label{accum}}
\end{figure}

To address this, when adjacent snapshots are very similar, FITrig can be applied after combining multiple snapshots, as illustrated in Fig. \ref{accum}. This approach ensures that the inputs fed into FITrig show noticeable variations. Snapshots can be combined progressively until there is enough difference between adjacent accumulated snapshots. The accumulation process preserves all information while increasing the contrast between the snapshots.

\subsubsection{Nyquist Sampling Effects on Detection Sensitivity}
\label{sec42}

With increasing $\Gamma$, the pulsar becomes relatively fainter and is more easily buried in noise. The sensitivity of the image-frequency domain FITrig is evaluated by the z-score of the pulsar tile, as a measure of the SNR. Tiles with z-scores above the threshold $\tau=6$ are forwarded for localisation. In this test, we employ the maximum spectral magnitude rather than harmonic summing to assess the detection robustness under conservative conditions, ensuring that our method remains effective even when harmonic power is negligible or obscured by noise.

Z-scores of the pulsar tile under different circumstances (pulsar frequency $f_p$, telescope arrays, $L$ and $\Gamma$) are shown in Table \ref{table423}. In it, $f_s$ is the sampling rate of the telescope array, and `-' means that the z-score of the pulsar tile under a certain combination of those circumstances is lower than $\tau=6$, indicating that the pulsar cannot be detected. According to the Nyquist-Shannon sampling theorem, when $f_s/(2\times f_p) \geq 1$, it is proper sampling; otherwise, it is under sampling.

From the table, as $\Gamma$ increases, the z-score of the pulsar tile decreases for the same $L$, telescope array, and $f_p$. This suggests that the pulsar becomes relatively fainter, making it harder to detect. When the length of the snapshot series ($L$) is longer, the $\eta_t$ of the pulsar tile becomes larger. However, $L$ does not have a significant effect on $z_t$, because the detection probability remains stable while longer series produce stronger spectral amplitudes. Thus, the detection will not be affected by $L$, as long as it is not too short.

Furthermore, when $f_s/(2\times f_p)$ is larger, the pulsar being detected can have a larger $\Gamma$, indicating that the pulsar is relatively fainter. Even if $f_s/(2\times f_p)$ is slightly smaller than 1, a pulsar with flux 20 times fainter than the surrounding steady features can be detected. In contrast, when $f_s/(2\times f_p)$ is much smaller than 1, where the pulsar signal is under sampled, only bright pulsars can be detected. 

\subsection{False Positive Rate}
\label{sec41}
\subsubsection{Image Domain FITrig}

By applying tLISI before the source finder SOFIA 2, not only is the computation sped up, but the false positive rate is also reduced. Using the simulated data, Table \ref{precision} shows the number of candidates detected by SOFIA 2 with and without tLISI under different $\Gamma$. For higher precision, $\tau$ is set to 6. The term ``\# Candidates'' indicates the number of output sources from SOFIA 2. A single pulsar is placed in the images. Two approaches are compared: ``FITrig'' refers to using tLISI to activate SOFIA 2 on relevant tiles, while ``SOFIA2'' refers to applying SOFIA 2 on the whole image.
\begin{table}[!t]
\centering
\begin{tabular}{ccccccccc}
\hline
\# Candidates &  & \multicolumn{7}{c}{\textbf{$\Gamma$}}\\
\textbf{Telescope} & \textbf{Approach} &\textbf{0.05} & \textbf{1} & \textbf{20} & \textbf{40} & \textbf{60} & \textbf{80} & \textbf{100}\\
\hline
\textbf{MeerKAT} & \textbf{FITrig} & 8 & 8 & 9 & 38 & 67 & 90 & 93\\
 & \textbf{SOFIA2} & 3989 & 3970 & 3868 & 3851 & 3846 & 3844 & 3841\\
\textbf{SKA AA2} & \textbf{FITrig} & 4 & 4 & 5 & 30 & 62 & 72 & 75 \\
 & \textbf{SOFIA2} & 3435 & 3428 & 3346 & 3339 & 3331 & 3329 & 3327 \\
\hline
\end{tabular}
\caption{Number of candidate transients in image domain detections.\label{precision}}
\end{table}

Currently, astronomers apply clustering algorithms to the candidates outputted by SOFIA 2 to determine whether they are true pulsars. As shown in the table, the number of candidate transients is greatly reduced (by a factor of up to 858.8) when using FITrig, making it easier for users to select the true pulsar(s) among the candidates.

\subsubsection{Image-Frequency Domain FITrig}

By image-frequency domain FITrig, the number of candidate pulsars after the interpolation is further reduced. With the threshold of $\tau=6$ and the sequence length of the snapshots of $L=1026$ (a reasonable choice to maintain both sensitivity and computational efficiency), the number of candidate transients corresponding to different circumstances is shown in Table \ref{freqprec}. Compared to the current source finder SOFIA 2, as shown in Table \ref{precision}, the number of false positives is greatly reduced.
\begin{table}[!t]
\centering
\begin{tabular}{ccccccccc}
\hline
\# Candidates &  & \multicolumn{7}{c}{\textbf{$\Gamma$}}\\
\textbf{Telescope} & \textbf{$f_s/(2\times f_p)$} &\textbf{0.05} & \textbf{1} & \textbf{20} & \textbf{40} & \textbf{60} & \textbf{80} & \textbf{100}\\
\hline
\textbf{MeerKAT} & \textbf{0.0982} & 1 & 1 & - & - & - & - & -\\
 & \textbf{0.9817} & 1 & 1 & 38 & - & - & - & -\\
  & \textbf{9.8175} & 1 & 1 & 14 & 34 & 38 & - & -\\
\textbf{SKA AA2} & \textbf{0.0982} & 1 & 1 & - & - & - & - & -\\
 & \textbf{0.9817} & 1 & 1 & 28 & - & - & - & -\\
  & \textbf{9.8175} & 1 & 1 & 1 & 16 & 25 & 31 & 32\\
\hline
\end{tabular}
\caption{Number of candidate transients localised by image-frequency domain FITrig.\label{freqprec}}
\end{table}

\subsubsection{False Positives for Non-Pulsar Snapshots}

Different sampling rates impact the number of false positives when images in a three-snapshot unit are all non-pulsar. The amount of false positives is influenced by the dirty beam associated with each telescope configuration.

The selection of pulsar candidates is based on statistical analysis using z-scores. As such, calculating z-scores solely on the tLISI matrix obtained from non-pulsar snapshots is not sensible. By calculating z-scores for tLISI of each tile, incorporating tiles from both pulsar and non-pulsar snapshots, we identify false positives in the non-pulsar snapshots. This is achieved by activating SOFIA 2 on tiles with z-score exceeding $\tau=6$. The number of false positives under various sampling periods ($1/f_s$, representing the sampling period) and the levels of pulsar faintness (represented by $\Gamma$) are shown in Tables \ref{fpaa2} and \ref{fpmeer} for SKA AA2 and MeerKAT telescope configurations, respectively.

According to the tables, brighter pulsars result in fewer false positives in non-pulsar snapshots. Compared to the results in Table \ref{precision}, the numbers of false positives in non-pulsar snapshots are generally lower and therefore acceptable, even when the time difference between adjacent snapshots extends to 16 seconds.

\section{Conclusions}
\label{seccon}

In this article, we develop FITrig, a GPU-accelerated technique, for ULP pulsar detection and localisation. A ULP pulsar typically has a period longer than twice the sampling period of the telescope. Unlike traditional methods that perform a thorough search across entire images, FITrig offers advantages by increasing sensitivity to faint pulsars, reducing susceptibility to false positives from noise, and delivering substantial gains in computational efficiency through massive parallelisation, with the improvement being especially obvious for large images. In this article, the effectiveness of FITrig is validated on a real dataset to demonstrate its practicality and applicability, and FITrig is then tested on simulated datasets to showcase its capability over existing methods in more challenging cases.

The key novelties of FITrig include:
\begin{itemize}
    \item \textbf{tLISI --- a novel transient-oriented IQA index:} A statistically-grounded IQA metric specially designed to highlight pulsar-specific variations through probability theory, enabling more reliable transient detection.
    \item \textbf{Elimination of deconvolution:} FITrig completely removes the need for deconvolution in the detection process while maintaining accuracy, significantly reducing computational overhead.
    \item \textbf{Dual-domain detection framework:} Two branches of FITrig (image domain and image-frequency domain approaches) are developed, significantly improving the capability to detect faint ULP pulsars.
    \item \textbf{GPU-powered real-time processing:} FITrig is highly parallelised with the GPU implementation, enabling real-time processing and scalability for large datasets.
\end{itemize}

Looking ahead, future research could explore whether filling the sampling gaps between observational snapshots could further enhance detection sensitivity. Machine learning methods, such as image completion \citep{SVT}, could be applied to complete the time-sequential pulsar images in the measurements. Furthermore, while this study focused on pulsar data (observed and simulated using radio telescope configurations), the application of FITrig may be expanded to a broader range of transients and wider observational bands in astronomy.

\section*{Acknowledgements}

The authors acknowledge support received from the STFC Grant (ST / W001969 / 1). The authors also acknowledge the use of the University of Oxford Advanced Research Computing (ARC; \citealt{arc}) facility in carrying out this work. Special thanks are due to Vlad Stolyarov for helpful discussion about fast imaging. Gratitude is extended to Fred Dulwich and Ben Mort for valuable guidance on OSKAR. We wish to express our sincere thanks to Mike Giles for helpful knowledge and practice about CUDA programming learnt from Course on CUDA Programming on NVIDIA GPUs (\url{https://people.maths.ox.ac.uk/gilesm/cuda/index.html}). Sincere appreciation is extended to Jack White for his helpful discussion regarding time-domain transient detection.

\bibliographystyle{elsarticle-harv} 
\bibliography{Bibliography.bib}

@article{CLEAN1,
	Author = {Hogbom, J.},
	Journal = {Astron. Astrophys. Suppl.},
	Number = {},
	Pages = {417-426},
	Title = "{Aperture synthesis with a non-regular distribution of interferometer baselines}",
	Volume = {15},
	Year = {1974}
}

@article{faintp,
    year = {2014},
    month = {},
    publisher = {},
    volume = {570},
    number = {A44},
    pages = {11},
    author = {X. Hou and D. A. Smith and L. Guillemot and others},
    title = {Six faint gamma-ray pulsars seen with the Fermi Large Area Telescope},
    journal = {A\&A}
}

@article{faintp2,
    year = {2021},
    month = {},
    publisher = {},
    volume = {64},
    number = {129562},
    pages = {},
    author = {Pei Wang and Di Li and Colin J. Clark and others},
    title = {FAST discovery of an extremely radio-faint millisecond pulsar from the Fermi-LAT unassociated source 3FGL J0318.1+0252},
    journal = {Science China Physics, Mechanics \& Astronomy}
}

@article{fdas1,
    year = {2002},
    month = {sep},
    publisher = {},
    volume = {124},
    number = {3},
    pages = {1788},
    author = {Scott M. Ransom and Stephen S. Eikenberry and John Middleditch},
    title = {Fourier Techniques for Very Long Astrophysical Time-Series Analysis},
    journal = {The Astronomical Journal}
}

@article{fdas2,
    year = {2018},
    month = {aug},
    publisher = {The American Astronomical Society},
    volume = {863},
    number = {1},
    pages = {L13},
    author = {Bridget C. Andersen and Scott M. Ransom},
    title = "{A Fourier domain `jerk' search for binary pulsars}",
    journal = {The Astrophysical Journal Letters}
}

@article{MEM10,
   Author = {Cornwell, T. and Evans, K.},
   Journal = {A\&A},
   Number = {},
   Pages = {77-83},
   Title = {A Simple Maximum Entropy Deconvolution Algorithm},
   Volume = {143},
   Year = {1985}
}

@article{MHD,
	author = {Urpin, V.},
	title = {Magnetohydrodynamic waves in the pulsar magnetosphere},
	journal = {A\&A},
	year = 2011,
	volume = 535,
	pages = "L5",
	month = "",
}

@article{MHD1,
    author = {Okamoto, Isao and Sigalo, Friday B.},
    title = "{Pulsar Magnetohydrodynamic Winds}",
    journal = {Publications of the Astronomical Society of Japan},
    volume = {58},
    number = {6},
    pages = {987-1013},
    year = {2006},
    month = {12}
}

@article{MHD2,
    author = {Tchekhovskoy, Alexander and Spitkovsky, Anatoly and Li, Jason G.},
    title = "{Time-dependent 3D magnetohydrodynamic pulsar magnetospheres: oblique rotators}",
    journal = {Monthly Notices of the Royal Astronomical Society: Letters},
    volume = {435},
    number = {1},
    pages = {L1-L5},
    year = {2013},
    month = {08},
}

@article{harmonic1,
    title = "{A novel greedy approach to harmonic summing using GPUs}",
    journal = {Astronomy and Computing},
    volume = {40},
    pages = {100621},
    year = {2022},
    author = {K. Adámek and J. Roy and W. Armour}
}

@article{iqara,
year = {2024},
volume = {274},
number = {2},
pages = {37},
author = {Xiaotong Li and Karel Adámek and Wesley Armour},
title = {Intensity-sensitive Quality Assessment of Extended Sources in Astronomical Images},
journal = {ApJS},
}

@article{m81,
author = {M. Bhardwaj and B. M. Gaensler and V. M. Kaspi and others},
year = {2021},
number = {L18},
title = "{A nearby repeating fast radio burst in the direction of M81}",
volume = {910},
journal = {ApJL}
}

@article{MKAT1,
author = {Booth, Roy and Jonas, Justin},
year = {2012},
pages = {101-104},
title = "{An overview of the MeerKAT project}",
volume = {16},
journal = {African Skies}
}

@article{MKAT3,
  author={Jonas, Justin},
  journal={Proc. IEEE}, 
  title="{MeerKAT --- The South African array with composite dishes and wide-band single pixel feeds}", 
  year={2009},
  volume={97},
  number={8},
  pages={1522-1530}
}

@article{realfast,
	Author = {C. J. Law and G. C. Bower and S. Burke-Spolaor and others},
	Journal = {ApJS},
	Number = {8},
	Pages = {},
	Title = "{realfast: Real-time, Commensal Fast Transient Surveys with the Very Large Array}",
	Volume = {236},
	Year = {2018}
}

@article{realpul1,
	Author = {Pelisoli, I. and Marsh, T.R. and Buckley, D.A.H. and others},
	Journal = {Nat Astron},
	Number = {},
	Pages = {931-942},
	Title = "{A 5.3-min-period pulsing white dwarf in a binary detected from radio to X-rays}",
	Volume = {7},
	Year = {2023}
}

@article{realpul2,
	Author = {Caleb, M. and Heywood, I. and Rajwade, K. and others},
	Journal = {Nat Astron},
	Number = {},
	Pages = {828-836},
	Title = "{Discovery of a radio-emitting neutron star with an ultra-long spin period of 76 s}",
	Volume = {6},
	Year = {2022}
}

@article{relativity,
	Author = {Hoffmann, B.},
	Journal = {Nature},
	Number = {},
	Pages = {667-668},
	Title = {Pulsars and a Possible New Test of General Relativity},
	Volume = {218},
	Year = {1968}
}

@article{relativity1,
author = {Gregory Desvignes and Michael Kramer and Kejia Lee and others},
title = {Radio emission from a pulsar’s magnetic pole revealed by general relativity},
journal = {Science},
volume = {365},
number = {6457},
pages = {1013-1017},
year = {2019}}

@article{scfinder,
    author = {Serra, Paolo and Oosterloo, Tom and Morganti, Raffaella and others},
    title = "{The ATLAS3D project – XIII. Mass and morphology of H i in early-type galaxies as a function of environment}",
    journal = {Monthly Notices of the Royal Astronomical Society},
    volume = {422},
    number = {3},
    pages = {1835-1862},
    year = {2012},
    month = {05},
}

@ARTICLE{sfind,
       author = {{Hopkins}, A.~M. and {Miller}, C.~J. and {Connolly}, A.~J. and others},
        title = {A New Source Detection Algorithm Using the False-Discovery Rate},
      journal = {AJ},
         year = 2002,
        month = feb,
       volume = {123},
       number = {2},
        pages = {1086-1094}
}

@article{sofia2,
    author = {Westmeier, T and Kitaeff, S and Pallot, D and others},
    title = "{sofia 2 - an automated, parallel H i source finding pipeline for the WALLABY survey}",
    journal = {Monthly Notices of the Royal Astronomical Society},
    volume = {506},
    number = {3},
    pages = {3962-3976},
    year = {2021},
    month = {07}
}

@article{sourcefinder1,
    author = {Hancock, P. J. and Murphy, T. and Gaensler, B. M. and others},
    title = "{Compact continuum source finding for next generation radio surveys}",
    journal = {Monthly Notices of the Royal Astronomical Society},
    volume = {422},
    number = {2},
    pages = {1812-1824},
    year = {2012},
    month = {04}
}

@article{srcdtc,
year = {2011},
month = {Mar},
publisher = {The American Astronomical Society},
volume = {731},
number = {2},
pages = {81},
author = {Cathryn M. Trott and Randall B. Wayth and Jean-Pierre R. Macquart and Steven J. Tingay},
title = "{Source detection in interferometric visibility data. I. Fundamental estimation limits}",
journal = {The Astrophysical Journal},
}

@article{SVT,
title = "{GPU accelerated singular value thresholding}",
journal = {SoftwareX},
volume = {23},
pages = {101500},
year = {2023},
author = {Xiaotong Li and Karel Adámek and Wes Armour}
}

@article{ulf,
    author = {Bezuidenhout, M. C. and Bhat, N. D. R. and Caleb, M. and others},
    title = {Slow and steady: long-term evolution of the 76-s pulsar J0901-4046},
    journal = {Monthly Notices of the Royal Astronomical Society},
    volume = {540},
    number = {3},
    pages = {2131-2145},
    year = {2025},
    month = {05},
    doi = {10.1093/mnras/staf773}
}

@article{WS1,
  author = {Offringa, A. R. and McKinley, B. and Hurley-Walker and others},
  title = {{WSClean: an implementation of a fast, generic wide-field imager for radio astronomy}},
  volume = {444},
  number = {1},
  pages = {606-619},
  year = {2014},
  journal = {MNRAS}
}

@book{RAbook,
    author    = "Burke, B. and Graham-Smith, F. and Wilkinson, P.",
    title     = "{An Introduction to Radio Astronomy}",
    year      = "2019",
    publisher = "Cambridge University Press",
    address   = "Cambridge",
    Edition = "4"
}

@book{RAbook2,
    author    = "Thompson, A. and Moran, J. and Swenson, G.",
    title     = "{Interferometry and Synthesis in Radio Astronomy}",
    year      = "2017",
    publisher = "Wiley",
    address   = "Oxford",
    Edition = "3"
}

@book{Pulsar,
  title = "{Handbook of Pulsar Astronomy}",
  author = {D. Lorimer and M. Kramer},
  year = {2004},
  month = {December},
  publisher = {Cambridge University Press},
  series = {Cambridge Observing Handbooks for Research Astronomers}
}

@INPROCEEDINGS{adass,
  author={Xiaotong Li and Karel Adámek and Wesley Armour},
  booktitle={Astronomical Data Analysis Software and Systems (ADASS) XXXIV}, 
  title="{GPU accelerated image quality assessment-based software for transient detection}", 
  year={2024},
  volume={},
  number={}}

@inproceedings{CUDA,
    author = {Kirk, David},
    title = "{NVIDIA CUDA software and GPU parallel computing architecture}",
    booktitle = {Proceedings of the 6th International Symposium on Memory Management},
    year = {2007},
    publisher = {},
    editor = {},
    pages = {103-104},
    volume = {7}
}

@inproceedings{MKAT2,  
    author={Davidson, David},
    booktitle="{ISAPE 2012}",
    title="{MeerKAT and SKA phase 1}",
    year={2012},
    volume={},
    number={},
    pages={1279-1282}
}

@inproceedings{Wider,
    author={Li, Xiaotong and Armour, Wesley},
    booktitle="{2022 26th International Conference on Pattern Recognition (ICPR)}", 
    title={Intensity-Sensitive Similarity Indexes for Image Quality Assessment}, 
    year={2022},
    volume={},
    number={},
    pages={1975-1981}
}

@misc{arc,
  title = "{University of Oxford Advanced Research Computing}",
  DOI = {10.5281/zenodo.22558},
  author = "Richards, A",
  year = 2015
}

@misc{OSKAR1,
  author = {Dulwich, Fred and Mort, Benjamin and Stolyarov, Vladislav and others},
  title = "{OxfordSKA/OSKAR}",
  DOI = {10.5281/zenodo.5722575},
  year = {2022}
}

@misc{ska1low,
  author = {P. Dewdney and J. Wagg and R. Braun and W. Turner},
  title = "{SKA1-LOW Configuration - Constraints \&
Performance Analysis}",
  howpublished ={\url{https://www.skao.int/sites/default/files/documents/d17-SKA-TEL-SKO-0000557_01_-DesignConstraints-1.pdf}},
  year = 2016
}

@ARTICLE{VLAapjs,
       author = {{Thompson}, A.~R. and {Clark}, B.~G. and {Wade}, C.~M. and {Napier}, P.~J.},
        title = "{The Very Large Array.}",
      journal = {The Astrophysical Journal Supplement Series},
         year = 1980,
        month = oct,
       volume = {44},
        pages = {151-167},
          doi = {10.1086/190688},
       adsurl = {https://ui.adsabs.harvard.edu/abs/1980ApJS...44..151T}
}

@misc{zenodoFIT,
    author={Li, Xiaotong},
    title={Fast Imaging Trigger v1.0.0},
    year={2024},
    DOI = {10.5281/zenodo.13901829}
}

@techreport{SKA1,
  author = "Dewdney, P. and Turner, W. and Millenaar, R. and others",
  title = "{SKA1 System Baseline Design}",
  institution = "SKA Organisation",
  address = "Macclesfield",
  Revision = "1",
  number = "SKA-TEL-SKO-DD-001",
  year = "2013"
}

@phdthesis{xiaotongthesis,
  author       = {Xiaotong Li}, 
  title        = {Image quality assessment-based fast imaging pipeline for transient detection with GPU acceleration},
  school       = {University of Oxford},
  year         = 2024,
  doi          = {10.5287/ora-g7nnok9o9}
}

\appendix
\clearpage
\begin{landscape}
\section{Appendix}
\label{app}

\begin{itemize}
    \item \textbf{Table \ref{abbrev}:} Abbreviations and definitions of technical terms used throughout the article.
    \item \textbf{Table \ref{table423}:} Sensitivity indicated by the z-score of the pulsar tile with the threshold of $\tau=6$, when applying FITrig to time sequential snapshots.
    \item \textbf{Table \ref{fpaa2}:} Number of false positives detected by FITrig in non-pulsar snapshots simulated using the SKA AA2 configuration.
    \item \textbf{Table \ref{fpmeer}:} Number of false positives detected by FITrig in non-pulsar snapshots simulated using the MeerKAT configuration.
\end{itemize}

\begin{table}[t]
\centering
\begin{tabular}{p{0.35\linewidth} p{0.12\linewidth} p{0.51\linewidth}}
\hline
Term & Abbreviation & Description \\
\hline

(Celestial) Transient & - &  A radio source whose emission varies noticeably over time in observations.\\
Compute Unified Device Architecture & CUDA & A parallel computing platform and programming model developed by NVIDIA. \\
Fast Imaging & FI & Technique that reconstructs images from visibilities collected over short timescales and identifies transient candidates through statistical analysis of these images. \\
Fast Imaging Trigger & FITrig & An image-based transient detection technique demonstrating high performance. \\
Image Quality Assessment & IQA & Metric to evaluate the quality of images. \\
Dirty Beam & - & The point spread function (PSF) of the telescope array resulting from its incomplete \textit{uv}-coverage.\\
Steady Features & - & The structures formed by steady sources together with the dirty beams convolved on them.\\
Steady Source & - & A radio source whose emission remains largely constant, showing no significant variability during observations.\\
Sky Brightness Distribution & SBD & Spatial distribution of brightness across the sky. \\
Transient-Oriented Low-Information Similarity Index & tLISI & Metric to quantify similarity in data with minimal information, focused on transient events. \\
Ultra-Long-Period Pulsars & ULP Pulsars & Pulsars with periods longer than twice the sampling period of the telescope. \\
\hline
\end{tabular}
\caption{Abbreviations and definitions of technical terms used throughout the article.}\label{abbrev}
\end{table}

\begin{table}[t]
\centering
\begin{tabular}{ccccccccccccccc}
\hline
\multicolumn{3}{c}{Z-score of pulsar tile} & \multicolumn{7}{c}{\textbf{$\Gamma$}}\\
\textbf{$f_s/(2\times f_p)$} & \textbf{Telescope} & \textbf{$L$} & \textbf{0.05} & \textbf{1} & \textbf{20} & \textbf{40} & \textbf{60} & \textbf{80} & \textbf{100}\\
\hline
\textbf{0.0982} & \textbf{MeerKAT} & \textbf{258} & 63.72 & 63.08 & - & - & - & - & -\\
 & & \textbf{514} & 63.76 & 63.08 & - & - & - & - & -\\
 & & \textbf{1026} & 63.77 & 63.23 & - & - & - & - & - \\
 & & \textbf{2050} & 63.81 & 63.42 & - & - & - & - & - \\
 & \textbf{SKA AA2} & \textbf{258} & 77.57 & 77.05 & - & - & - & - & - \\
 & & \textbf{514} & 77.53 & 77.26 & - & - & - & - & - \\
 & & \textbf{1026} & 77.49 & 77.04 & - & - & - & - & - \\
 & & \textbf{2050} & 77.47 & 76.76 & - & - & - & - & - \\
\textbf{0.9817} & \textbf{MeerKAT} & \textbf{258} & 63.52 & 63.68 & 7.79 & - & - & - & - \\
 & & \textbf{514} & 63.53 & 63.72 & 7.92 & - & - & - & - \\
 & & \textbf{1026} & 63.54 & 63.79 & 8.64 & - & - & - & - \\
 & & \textbf{2050} & 63.57 & 63.94 & 8.79 & - & - & - & - \\
 & \textbf{SKA AA2} & \textbf{258} & 76.49 & 76.48 & 13.07 & - & - & - & - \\
 & & \textbf{514} & 76.49 & 76.46 & 15.32 & - & - & - & - \\
 & & \textbf{1026} & 76.45 & 76.29 & 18.81 & - & - & - & - \\
 & & \textbf{2050} & 76.44 & 75.97 & 15.76 & - & - & - & - \\
\textbf{9.8175} & \textbf{MeerKAT} & \textbf{258} & 63.43 & 63.61 & 52.53 & 21.15 & 9.59 & - & - \\
 & & \textbf{514} & 63.43 & 63.60 & 52.54 & 21.10 & 9.50 & - & - \\
 & & \textbf{1026} & 63.44 & 63.62 & 53.15 & 21.79 & 9.84 & - & - \\
 & & \textbf{2050} & 63.46 & 63.65 & 51.95 & 20.13 & 8.97 & - & - \\
 & \textbf{SKA AA2} & \textbf{258} & 76.22 & 76.88 & 69.43 & 35.87 & 17.48 & 9.94 & 6.31 \\
 & & \textbf{514} & 76.21 & 76.85 & 71.00 & 41.00 & 20.73 & 11.89 & 7.59\\
 & & \textbf{1026} & 76.20 & 76.80 & 71.29 & 43.62 & 22.70 & 13.17 & 8.46 \\
 & & \textbf{2050} & 76.21 & 76.72 & 70.46 & 42.94 & 22.33 & 12.98 & 8.35\\
\hline
\end{tabular}
\caption{Sensitivity indicated by the z-score of the pulsar tile with the threshold of $\tau=6$, when applying FITrig to time sequential snapshots.\label{table423}}
\end{table}

\begin{table}[!t]
\centering
\begin{tabular}{ccccccccccccccccc}
\hline
\# False positives & \multicolumn{16}{c}{\textbf{$1/f_s$ (sec)}}\\ \textbf{$\Gamma$} &
\textbf{1} & \textbf{2} &\textbf{3} & \textbf{4} & \textbf{5} & \textbf{6} & \textbf{7} & \textbf{8} & \textbf{9} & \textbf{10} & \textbf{11} & \textbf{12} & \textbf{13} & \textbf{14} & \textbf{15} & \textbf{16}\\
\hline
\textbf{0.05} & 0 & 0 & 0 & 0 & 0 & 0 & 0 & 0 & 0 & 0 & 0 & 0 & 0 & 0 & 0 & 0\\
\textbf{1} & 0 & 0 & 0 & 0 & 0 & 0 & 0 & 0 & 0 & 0 & 0 & 0 & 0 & 0 & 0 & 0\\
\textbf{20} & 3 & 1 & 2 & 6 & 2 & 9 & 12 & 9 & 19 & 27 & 24 & 26 & 21 & 19 & 25 & 21\\
\textbf{40} & 15 & 16 & 6 & 17 & 11 & 16 & 22 & 23 & 24 & 33 & 26 & 29 & 26 & 22 & 27 & 27\\
\textbf{60} & 19 & 23 & 10 & 18 & 13 & 16 & 23 & 23 & 27 & 33 & 29 & 29 & 27 & 22 & 28 & 27\\
\textbf{80} & 19 & 26 & 10 & 18 & 13 & 17 & 23 & 25 & 27 & 33 & 29 & 29 & 27 & 22 & 28 & 27\\
\textbf{100} & 20 & 26 & 10 & 20 & 13 & 17 & 23 & 25 & 27 & 33 & 29 & 29 & 27 & 22 & 28 & 27\\
\hline
\end{tabular}
\caption{Number of false positives detected by FITrig in non-pulsar snapshots simulated using the SKA AA2 configuration.\label{fpaa2}}
\end{table}

\begin{table}[!t]
\centering
\begin{tabular}{ccccccccccccccccc}
\hline
\# False positives & \multicolumn{16}{c}{\textbf{$1/f_s$ (sec)}}\\ \textbf{$\Gamma$} &
\textbf{1} & \textbf{2} &\textbf{3} & \textbf{4} & \textbf{5} & \textbf{6} & \textbf{7} & \textbf{8} & \textbf{9} & \textbf{10} & \textbf{11} & \textbf{12} & \textbf{13} & \textbf{14} & \textbf{15} & \textbf{16}\\
\hline
\textbf{0.05} & 0 & 0 & 0 & 0 & 0 & 0 & 0 & 0 & 0 & 0 & 0 & 0 & 0 & 0 & 0 & 0\\
\textbf{1} & 0 & 0 & 0 & 0 & 0 & 0 & 0 & 0 & 0 & 0 & 0 & 0 & 0 & 0 & 0 & 0\\
\textbf{20} & 1 & 1 & 0 & 3 & 0 & 3 & 3 & 1 & 4 & 12 & 4 & 13 & 5 & 14 & 16 & 11\\
\textbf{40} & 20 & 17 & 0 & 20 & 23 & 14 & 18 & 14 & 30 & 28 & 28 & 35 & 27 & 28 & 37 & 39\\
\textbf{60} & 34 & 30 & 0 & 27 & 31 & 19 & 25 & 25 & 37 & 32 & 30 & 40 & 34 & 30 & 41 & 43\\
\textbf{80} & 35 & 34 & 12 & 27 & 35 & 21 & 27 & 26 & 39 & 33 & 30 & 40 & 34 & 31 & 42 & 44\\
\textbf{100} & 38 & 34 & 20 & 28 & 36 & 24 & 27 & 26 & 40 & 33 & 31 & 40 & 35 & 31 & 42 & 44\\
\hline
\end{tabular}
\caption{Number of false positives detected by FITrig in non-pulsar snapshots simulated using the MeerKAT configuration.\label{fpmeer}}
\end{table}

\end{landscape}

\end{document}